\documentclass[%
reprint,
superscriptaddress,
amsmath,amssymb,
aps,
floatfix,
]{revtex4-2}

\usepackage{graphicx}
\usepackage{dcolumn}
\usepackage{bm}
\usepackage{hyperref}
\usepackage[dvipsnames]{xcolor}
\usepackage{color}
\hypersetup{colorlinks = true,
urlcolor  = black, 
linkcolor = NavyBlue, 
citecolor = ForestGreen}

\usepackage{amsthm}

\newcommand{\nbr}[1]{\mathcal{N}(#1)}
\newcommand{\nE}{\mathrm{e}}
\newcommand{\D}{\mathrm{d}}
\newcommand{\A}{\langle \mathcal{A} \rangle_\tau}
\newcommand{\AD}{\langle \mathcal{A}^\mathrm{D} \rangle_\kappa}
\newcommand{\PS}{\mathrm{PS}}

\newcommand{\eqL}[1]{\label{eq:#1}}
\newcommand{\eqR}[1]{Eq.~(#1)}

\newcommand{\figL}[1]{\label{fig:#1}}
\newcommand{\figR}[1]{Fig.~(#1)}

\newcommand{\appL}[1]{\label{app:#1}}

\newtheorem{proposition}{Proposition}

\begin{document}
\title{Bound on annealing performance from stochastic thermodynamics, with application to simulated annealing}

\author{Yutong Luo}
\email{yutong.luo21@imperial.ac.uk}
\affiliation{%
 Blackett Laboratory, Imperial College London, London SW7 2AZ, United Kingdom
}%
\affiliation{Department of Physics, Southern University of Science and Technology, Shenzhen 518055, China}

\author{Yi-Zheng Zhen}
\email{zhenyizheng@ustc.edu.cn}
\affiliation{Hefei National Research Center for Physical Sciences at the Microscale and School of Physical Sciences, University of Science and Technology of China, Hefei 230026, China}
\affiliation{Shanghai Research Center for Quantum Science and CAS Center for Excellence in Quantum Information and Quantum Physics, University of Science and Technology of China, Shanghai 201315, China}

\author{Xiangjing Liu}%
\email{liuxj@mail.bnu.edu.cn}
\affiliation{Department of Physics, Southern University of Science and Technology, Shenzhen 518055, China}

\author{Daniel Ebler}%
\email{ebler.daniel1@huawei.com}
\affiliation{Theory Lab, Central Research Institute, 2012 Labs, Huawei Technology Co. Ltd., Hong Kong Science Park, Hong Kong SAR, China}
\affiliation{Department of Computer Science, The University of Hong Kong, Pokfulam Road, Hong Kong SAR, China}

\author{Oscar Dahlsten}%
\email{oscar.dahlsten@cityu.edu.hk}
\affiliation{Department of Physics, City University of Hong Kong, Tat Chee Avenue, Kowloon, Hong Kong SAR, China
}%
\affiliation{Shenzhen Institute for Quantum Science and Engineering and Department of Physics,
Southern University of Science and Technology, Shenzhen 518055, China}
\affiliation{Institute of Nanoscience and Applications, Southern University of Science and Technology, Shenzhen 518055, China}

\date{\today}

\begin{abstract} Annealing is the process of gradually lowering the temperature of a system to guide it towards its lowest energy states. In an accompanying paper [Luo \textit{et al.} \href{https://link.aps.org/doi/10.1103/PhysRevE.108.L052105}{Phys. Rev. E 108, L052105 (2023)}], we derived a general bound on annealing performance by connecting annealing with stochastic thermodynamics tools, including a speed-limit on state transformation from entropy production. We here describe the derivation of the general bound in detail. In addition, we analyze the case of simulated annealing with Glauber dynamics in depth. We show how to bound the two case-specific quantities appearing in the bound, namely the activity, a measure of the number of microstate jumps, and the change in relative entropy between the state and the instantaneous thermal state, which is due to temperature variation. We exemplify the arguments by numerical simulations on the SK model of spin-glasses. 
\end{abstract}

\maketitle


\section{Introduction}
Simulated annealing (SA) is a heuristic optimization algorithm to approximate the global minimum of a function in a large search space~\cite{Kirkpatrick1983, Bertsimas1993, Ingber1993, salamon2002facts}.
The algorithm models the physical annealing process in metallurgy~\cite{callister2018materials}: by treating the function to be optimized as the energy landscape of a system, the lowest energy state corresponding to the global minimum can be evolved via a gradually cooling process.
SA has been massively used in a wide variety of real applications~\cite{BERNARDI2015872, EKREN2010592, MEIRI2006842} and has stimulated novel development in hardware~\cite{Cai2020,Mohseni2022}.

The ideal SA finds the global optimum of any function from a quasistatic cooling process, requiring infinite annealing time~\cite{Hajek1988}. However, in a realistic setting, the cooling schedule is performed in a finite time. As a result, real SA algorithms do not guarantee the global optimum but only find it probabilistically.
Significant efforts have been made to investigate the convergence of real SA algorithms~\cite{Romeo1991,Andresen1996,Desai1999}, to analyze the effect of finite-length cooling schedules~\cite{Strenski1991,nourani1998comparison,Nolte2000,Fontenas2003}, and to gauge the performance of SA in specific problems~\cite{Albrecht2004,Nolte1996}.
In an accompanying paper~\cite{short}, we have derived an analytical bound on the performance of annealing by using recent methods from finite-time stochastic thermodynamics~\cite{Shiraishi2018,Dechant2021,Tan2022, Parrondo_2009,Esposito2010,Esposito2011, Zhen2021,Schmiedl2007}. The resulting bound holds for all cooling schedules. 

In this paper, we provide the full details of the results in Ref.~\cite{short} and show explicitly how the theorems and methods in finite-time stochastic thermodynamics can be adapted to quantify the SA performance. We derive a method for bounding the activity in SA, a measure of the number of microstate jumps. We also show how to bound the change in relative entropy between the state and the instantaneous thermal state which is due to temperature variation. Our arguments involve two conjectures which we describe and justify, and which we hope may spark further research. 

The paper is organized as follows. In Sec.~\ref{sec:StochasticBasics}, we introduce key concepts in stochastic thermodynamics that we shall use. In Sec.~\ref{sec:GeneralBound}, we prove the bound proposed in Ref.~\cite{short} which characterizes the performance of any (physical or simulated) annealing process. In Sec.~\ref{sec:Application2SA}, we apply this bound to SA and discuss in detail the assessment of SA performance from accessible parameters. Finally, in Sec.~\ref{sec:Conclusion}, we conclude our results.

\section{Stochastic thermodynamics}\label{sec:StochasticBasics}

We start by briefly introducing the stochastic thermodynamical framework of annealing processes, using standard terminology of e.g.\ Refs.~\cite{Esposito2010,Esposito2011,Dechant2021}.

There is a system of interest. The system has $N$ energy levels denoted as $\{E_i\}_{i=1}^N$. The statistical state of the system is defined as a vector $p(t) = \left[p_1(t),\dots,p_N(t)\right]$, where $p_i(t)$ is the probability of the system on the energy level $E_i$ at time $t$. The average energy of the system is
\begin{equation}
E_p:=\sum_{i=1}^N p_iE_i.
\end{equation}

In annealing, the system of interest is in contact with
a heat bath whose temperature $T$ can be controlled to vary in time. 
If the system is not in thermal equilibrium, it will then thermalize, that is, the state of the system  will evolve towards the thermal Gibbs' state $\gamma = \left[\gamma_i,\dots,\gamma_N\right]$,
with $\gamma_i = \exp\left(-\beta E_i\right)/Z$, where $Z = \sum_i\exp\left(-\beta E_i\right)$ is the partition function and $\beta = 1/T$ is the inverse temperature (The Boltzmann constant $k_\mathrm{B}$ is taken to be 1). In annealing, the thermal state $\gamma(t)$ is time-dependent due to the varying temperature $T(t)$.

The dynamics of a thermalization process are described by a master equation~\cite{Kampen2007,Esposito2010,levin2017markov}:
\begin{equation}\label{eq:master_equation}
    \dot{p}_i(t) 
    =\sum_{j(\neq i)}\Gamma_{ij}(t)p_j(t)-\Gamma_{ji}(t)p_i(t),
\end{equation}
where $\Gamma_{ij}(t)$ is the generator satisfying $\sum_i\Gamma_{ij}(t) = 0, \forall i$ and $\Gamma_{ij}(t) \ge 0, \forall i\neq j$. In addition, we suppose that the detailed balance condition is satisfied in the process, i.e.,
\begin{equation}
    \Gamma_{ij}(t)\gamma_j(t) = \Gamma_{ji}(t)\gamma_i(t),\quad \forall i,j.,
    \eqL{detailed_balance}
\end{equation}
which is commonly assumed in stochastic thermodynamics (See e.g.\ Ref.~\cite{Esposito2011,Shiraishi2018,Dechant2021}). 

As the system evolves, there are changes in its energy and entropy. It is common to break the change in (average) energy $E_p$ into two parts:
\begin{align}
dE_p & =d\left( \sum_i p_iE_i \right)=\sum_i E_idp_i+\sum_i p_idE_i \nonumber\\
&:=dQ+dW, 
\end{align}
where $Q$ is called the heat added to the system and $W$ is called the work done to the system. In annealing, the energy spectrum is unchanged and $dE_p=dQ$. The thermodynamic entropy of the system (again for $k_\mathrm{B}=1$), 
\begin{equation}
S_p:=-\sum_i p_i\ln p_i,    
\end{equation}
also changes due to the evolution of $p$. Note that we shall make no distinctions between the thermodynamic entropy and Shannon entropy since the two quantities are the same within our setting. For the thermal state $\gamma$ transitioning to a new thermal state at an infinitesimally different temperature, as in quasistatic annealing, one can show $\beta dQ=dS_p$. If the state is not a thermal state, one has $dS_p\geq \beta dQ$. The quantity 
\begin{equation}\label{eq: entroprodrate}
\dot{\Sigma}(t) = \dot{S}_p(t) - \beta(t)\dot{Q}(t),
\end{equation}
is thus non-negative in the case where the system is not in thermal equilibrium. $\dot{\Sigma}$ is commonly termed the {\em entropy production} rate. The entropy production when two systems 1 and 2 (like the system and the bath) interact is by definition $dS_1+dS_2$, in line with early ideas of entropy being akin to a fluid existing in and flowing between different systems and possibly being created~\cite{prigogine1963introduction}. Here the bath is in a thermal state, such that $dS_1+dS_2=dS_1+\beta(t)dQ_2=dS_1-\beta(t)dQ_1$, justifying the name {\em entropy production} rate for $\dot{\Sigma}$ (see e.g.~Ref.~\cite{Esposito2011} for more).

To characterize the performance of an annealing protocol run in finite time $\tau$, we consider the 1-norm distance ~\cite{Shiraishi2018,Dechant2021,Tan2022} to the final thermal state $\gamma(\tau)$:
\begin{equation}
        L_t := \sum_i|p_i(t)-\gamma_i(\tau)|.
        \eqL{L_t}
\end{equation}
The probability of not having a globally optimal state at time $\tau$ when $T(\tau)=0$, $p(\text{non-optimal})$ is bounded by $L_\tau$. To show this, notice that
the 1-norm distance is twice the total variation distance~\cite{levin2017markov} such that 
\begin{align}
    \frac{1}{2}L_\tau&=\max_{\mathrm{events}} |p(\mathrm{events})-\gamma(\mathrm{events})|\nonumber\\
    &\geq |p(\text{non-optimal})-\gamma(\text{non-optimal})|\nonumber\\
    &=p(\text{non-optimal}),
\end{align}
where in the last line we used the fact that $\gamma(\text{non-optimal}) = 0$ at zero temperature.
$L_\tau$ thus can be considered as the performance error of the annealing. In the next section,
an upper bound of $L_\tau$ in terms of $L_0$ for a general annealing process will be given.

\section{General bound on annealing in finite time}\label{sec:GeneralBound}
We then consider such an annealing process, where
a system initially in thermal equilibrium with a heat bath and the bath temperature is continuously decreased from $T_{\mathrm{i}}=T(0)$ to $T_{\mathrm{f}}=T(\tau)$ within a finite time $\tau$. In the accompanying paper~\cite{short}, we derived a general limit on the annealing performance error represented by $L_\tau$. Here, we provide the details of the derivation in the following three subsections.

\subsection{Splitting the relative entropy}
A critical challenge in evaluating the effectiveness of an annealing process is to measure the deviation of the system state, $p(\tau)$, from the reference thermal state $\gamma(\tau)$. While the 1-norm distance, $L_\tau$, is a suitable metric for this purpose, its direct calculation can be difficult. However, relative entropy, a comparison tool widely adopted in information theory to compare two probability distributions~\cite{Cover1991}, proves to be a good alternative to start with, due to its deep connection to non-equilibrium thermodynamics.
The relative entropy between probability distributions $p(t)$ and $\gamma(t)$ is defined as
\begin{equation}
    S(p(t)||\gamma(t)) = \sum_i p_i(t)\ln\dfrac{p_i(t)}{\gamma_i(t)}.
\end{equation}
Its close relationship with free energy is apparent via the expression,
\begin{equation}
    F(t)-F_{eq}(t)=T(t)S(p(t)||\gamma(t)),
\end{equation}
where the non-equilibrium free energy $F(t):=E_p(t)-T(t)S_p(t)$ and the equilibrium free energy $F_{eq}(t):=E_\gamma(t)-T(t)S_\gamma(t)$ with $E_\gamma(t)$ and $S_\gamma(t)$ being the energy and entropy of the thermal state $\gamma(t)$, respectively.

Taking the time derivative of $S(p||\gamma)$, one gets
\begin{align}
    \dfrac{\mathrm{d}}{\mathrm{d}t}S(p||\gamma)
    &= \dot{p}\frac{\partial } {\partial p}S(p||\gamma)+\dot{\gamma}\frac{\partial}{ \partial \gamma}S(p||\gamma) \nonumber\\
    &= \dot{p}\frac{\partial } {\partial p}S(p||\gamma)+\dot{\beta}\frac{\partial}{ \partial \beta}S(p||\gamma) \nonumber\\
    &= -\dot{\Sigma} + \dot{\mathcal{I}}, 
    \eqL{relative_entropy_dot}
\end{align}
We immediately see that a new rate,  $\dot{\mathcal{I}}:=\dot{\beta}\frac{\partial}{ \partial \beta}S(p||\gamma) $, compared to a fixed temperature scenario,  appears at the end of \eqR{\ref{eq:relative_entropy_dot}}. $\mathcal{I}$ is an annealing-related quantity, arising from the temperature variation. A direct calculation shows that 
\begin{align}
\dot{\mathcal{I}} = \dot{\gamma}\frac{\partial}{ \partial \gamma}S(p||\gamma) =- \sum_i\dot{\gamma}_i\dfrac{p_i}{\gamma_i}
= (E_p - E_\gamma)\dot{\beta},
\end{align}
where the expression  
\begin{equation}
    \dot{\gamma}_i     = \dot{\beta}(E_\gamma - E_i)\gamma_i,
    \eqL{gamma_dot}
\end{equation} is used in the last equality.

We now comment on the physical meaning of these terms from~\eqR{\ref{eq:relative_entropy_dot}}. From Eq.~\eqref{eq: entroprodrate},
the entropy production rate, $\dot{\Sigma}$, is non-zero when the system thermalizes towards a fixed thermal state. We imagine that, in ideal annealing processes, the change of the thermal states can be regarded as the switching of the system to heat baths at different temperatures. The disconnection and re-connection of the system with baths should not alter the system state instantaneously, resulting in no entropy production during the change of the reference thermal state. However, the variation of the reference state does affect the value of the relative entropy. This contribution is represented by $\dot{\mathcal{I}}$. 

We proceed to split the relative entropy. Integrating both sides of \eqR{\ref{eq:relative_entropy_dot}} from 0 to $\tau$ gives us 
\begin{equation}
    S(p(\tau)||\gamma(\tau)) - S(p(0)||\gamma(0)) = -\Sigma(\tau) + \mathcal{I}(\tau),
\end{equation}
where $\Sigma(\tau) = \int_0^\tau \dot{\Sigma}\mathrm{d} t$ and 
\begin{equation}
    \mathcal{I}(\tau) = \int_0^\tau (E_p-E_\gamma)\dot{\beta}\mathrm{d} t.
    \label{eq:I_tau}
\end{equation}

Since the system is assumed to be in equilibrium initially, which is the most common case in annealing, we have $S(p(0)||\gamma(0)) = 0$ and therefore obtain,
\begin{equation}
    S(p(\tau)||\gamma(\tau)) = -\Sigma(\tau) + \mathcal{I}(\tau).
    \label{eq:relative_entropy}
\end{equation}

Using Pinsker's inequality~\cite{pinsker1964information}, $L_\tau^2\le 2S(p(\tau)||\gamma(\tau))$, we arrive at an upper bound on $L_\tau^2$,
\begin{equation}
    L_\tau^2 \le 2\left[-\Sigma(\tau) + \mathcal{I}(\tau)\right].
    \label{eq:L^2_bound}
\end{equation}

\subsection{Bounding entropy production during the annealing}
\label{sec:speed_limit}
In this subsection, we, inspired by Ref.~\cite{Shiraishi2018}, derive a lower bound for the entropy production $\Sigma(\tau)$ in the varying temperature case.
For the clarity of equations, we omit the time dependence of  $p_i(t)$ and $\Gamma_{ij}(t)$. 

First, we rewrite $\dot{\Sigma}(t)$ as 
\begin{align}
    \dot{\Sigma}(t) 
    &=  -\sum_i \dot{p}_i\left(\ln p_i + \beta(t) E_i\right) \nonumber\\
    &= \dfrac{1}{2} \sum_{i\neq j}\left(\Gamma_{ij}p_j - \Gamma_{ji}p_i\right)\ln\frac{\Gamma_{ij}p_j}{\Gamma_{ji}p_i} \nonumber\\
    &\ge \sum_{i\neq j} \dfrac{\left(\Gamma_{ij}p_j - \Gamma_{ji}p_i\right)^2}{\Gamma_{ij}p_j + \Gamma_{ji}p_i},
    \label{eq:entropy_prod_rate}
\end{align}
where we have used Eq.~\eqref{eq:master_equation}, the relation $\beta(t)E_i=-\ln \gamma_i-\ln Z $ in the second line, and the inequality $(x-y)\ln (x/y) \geq 2(x-y)^2/(x+y)$ in the third line, as suggested by Ref.~\cite{Shiraishi2018}.

Meanwhile, using the Cauchy-Schwarz inequality, we have
\begin{align}
    \sum_i |\dot{p}_i| 
    &= \sum_{i\neq j}\left|\Gamma_{ij}p_j - \Gamma_{ji}p_i\right| \nonumber\\
    &\le \sqrt{\left(\sum_{i\neq j} \dfrac{\left(\Gamma_{ij}p_j - \Gamma_{ji}p_i\right)^2}{\Gamma_{ij}p_j + \Gamma_{ji}p_i}\right)\left(\sum_{i\neq j}\Gamma_{ij}p_j + \Gamma_{ji}p_i\right)} \nonumber\\
    &\le \sqrt{2\dot{\Sigma}(t)\mathcal{A}(t)},
\end{align}
where we have defined the activity
\begin{equation}
    \mathcal{A}(t) := \sum_{i}\sum_{j(\neq i)}\Gamma_{ij}(t)p_j(t),
    \label{eq:activity}
\end{equation}
which is the expected jumping rate among different states at time $t$. We denote the expected number of jumps during the time interval $[0,\tau]$ as
\begin{align}\label{eq: expectN}
\langle N_{\mathrm{jumps}}\rangle =\int_0^\tau   \mathcal{A}(t) dt = \A\tau,
\end{align}
where $\A:=\frac{1}{\tau}\int_0^\tau\mathcal{A}(t) dt$. (See also discussion around \eqR{\ref{eq:Njumpsdiscrete}} for further explanation of why we call this the expected number of jumps).

Then, the 1-norm distance between $p(0)$ and $p(\tau)$ satisfies
    \begin{align}
        \sum_i|p_i(0)-p_i(\tau)| 
        &\le \sum_i\int_0^\tau \D t |\dot{p}_i(t)|  \nonumber\\
        &\le \int_0^\tau \D t  \sqrt{2\dot{\Sigma}(t)\mathcal{A}(t)} \nonumber\\
        &\le\sqrt{2\Sigma(\tau)\A\tau},
        \label{eq:speed_limit_mid_step}
    \end{align}
Equivalently, we obtain
\begin{equation}
    \Sigma(\tau) \ge \dfrac{\left(\sum_i|p_i(0)-p_i(\tau)|\right)^2}{2\A\tau}.
    \label{speed_limit_origin}
\end{equation}
That lower bound on $\Sigma(\tau)$ is termed the {\em speed limit} from entropy production on the transformation from the initial state $p(0)$ to the final $p(\tau)$.

As a concrete example, consider a  quasistatic process, which has zero entropy production ($\Sigma(\tau)=0$). Then the speed limit implies that for a finite activity $\A$, a transition between two distinct statistical states requires $\tau \rightarrow \infty$.

Since $\sum_i|p_i(0)-p_i(\tau)|$ is a distance, by triangular inequality, we have
    \begin{align}\label{eq: systdist}
        &\sum_i|p_i(0)-p_i(\tau)| \nonumber \\ 
        &\ge \Big|\sum_i|p_i(0)-\gamma_i(\tau)|-\sum_i|p_i(\tau)-\gamma_i(\tau)|\Big| \nonumber\\
        &= |L_0-L_\tau|.
    \end{align}
Substituting Eq.~\eqref{eq: systdist} into \eqR{\ref{speed_limit_origin}}, we finally obtain a lower bound of the entropy production for a general annealing process.
\begin{equation}
    \Sigma(\tau) \ge \dfrac{(L_0-L_\tau)^2}{2\A\tau}.
    \label{eq:speed_limit}
\end{equation}

\subsection{General bound}

Combining Eqs.~\eqref{eq:L^2_bound} and \eqref{eq:speed_limit}, we have
\begin{equation}\label{eq:UQineq}
    L_\tau^2 \le - \dfrac{(L_0-L_\tau)^2}{\A\tau} + 2\mathcal{I}(\tau).
\end{equation}
Eq.~\eqref{eq:UQineq} can be regarded as a unary quadratic inequality for $L_\tau$.  Solving this inequality for $L_\tau$ gives us
\begin{equation}
    L_\tau \le \dfrac{L_0 + \sqrt{\langle\mathcal{A}\rangle_\tau\tau\left(-L_0^2 + 2\mathcal{I}(\tau)\left(\langle\mathcal{A}\rangle_\tau\tau + 1\right)\right)}}{\langle\mathcal{A}\rangle_\tau\tau+1}.
    \label{eq:main_bound}
\end{equation}

This is the general bound of an annealing process as obtained in Ref.~\cite{short}. It is worth noticing that \eqR{\ref{eq:main_bound}} is applicable for any system energy landscape and any cooling schedule as long as the thermalization dynamics respect the detailed balance condition in \eqR{\ref{eq:detailed_balance}}. 
Since the terms inside the square root in \eqR{\ref{eq:main_bound}} are non-negative given \eqR{\ref{eq:UQineq}}, $\A$ and $\mathcal{I}(\tau)$ obey a trade-off relation:
\begin{equation}
    \mathcal{I}(\tau) \ge \dfrac{L_0^2}{2\left(\A\tau+1\right)}.
    \label{eq:trade_off}
\end{equation}
This can be understood as a `speed limit' related to $\mathcal{I}(\tau)$, which can only be small if the expected number of jumps $\langle N_{\mathrm{jumps}}\rangle = \A\tau$ is large, given that $L_0$ is fixed.

In the quasistatic limit associated with the protocol duration $\tau\rightarrow\infty$, the tightness of our bound [\eqR{\ref{eq:main_bound}}] for the error $L_\tau$ is guaranteed. This can be seen by investigating how our bound on SA performance scales with $\tau$. Let the rate of varying temperature $\dot{T}$ decrease with $\tau$. In this case, $\mathcal{I}(\tau)$ will diminish as $\tau$ increases. This can be seen from bounding the expression of $\mathcal{I}(\tau)$ in Eq.~(\ref{eq:I_tau}): 
\begin{equation}
    \mathcal{I}(\tau) = \int_0^\tau (E_p-E_\gamma)\dot{\beta}\mathrm{d} t \le \max_{t\in[0,\tau]} (E_p-E_\gamma)\Delta\beta, 
\end{equation}
with $\Delta\beta\equiv \beta_\mathrm{f} - \beta_\mathrm{i}$ being the maximal range of the inverse temperature. As $\tau$ increases, the thermal state of the system changes slower, leading to a smaller $\max_{t\in[0,\tau]} (E_p-E_\gamma)$ that measures the departure from the equilibrium energy. We can define the downward scaling of $\mathcal{I}(\tau)$ for large $\tau$ in terms of the big O notation for an asymptotic upper bound: $\mathcal{I}(\tau) = \mathcal{O}(\tau^{-\alpha})$, where $0<\alpha\le 1$ is some scalar. An upper bound on $\alpha$ is in fact imposed by the trade-off relation [Eq.~(\ref{eq:trade_off})] (assuming $\A$ is finite (bounded)). Substituting that upper bound of $\mathcal{I}(\tau)$ into \eqR{\ref{eq:main_bound}}, the upper bound on $L_\tau$, we obtain $L_\tau = \mathcal{O}(\tau^{-\alpha/2})$. This scaling implies the tightness of the bound for quasistatic protocols: when $\tau\rightarrow\infty$, the protocol approaches being quasistatic and both $L_\tau$ and its bound vanish.

We now further analyse how to evaluate \eqR{\ref{eq:main_bound}} when using it to bound the performance of SA, in particular, how to evaluate
 $\A$ and $\mathcal{I}(\tau)$ without calculating the evolution of the statistical state of the system, $p(t)$. 

\section{Application to SA}\label{sec:Application2SA}
In this section we apply the bound on annealing performance, \eqR{\ref{eq:main_bound}}, to the case of SA.

SA is an optimization algorithm inspired by real annealing processes. The algorithm treats the cost function as the energy of a system and uses a control parameter called `temperature' to anneal it. The physical principle is that a system at thermal equilibrium has a higher probability of staying in its ground state at lower temperatures. The algorithm seeks to find the minimal cost by gradually lowering the temperature and moving closer to the final thermal equilibrium. Therefore, the success of the optimization depends on how far the system is from this equilibrium state.

In SA, the annealing time $\tau$ used in the previous section is replaced by the discrete time steps $\kappa$. $L_\kappa \equiv \sum_i |p_i(\kappa) - \gamma_i(\kappa)|$ becomes the error from the target state.  As we will show, a discrete-time analogue of \eqR{\ref{eq:main_bound}} provides an upper bound to the error when the SA is run in $\kappa$ steps.

To illustrate how to evaluate $\AD$ and $\mathcal{I}(\kappa)$, the discrete version of $\A$ and $\mathcal{I}(\tau)$, without referring to the intermediate statistical state of the system, we first briefly set up the framework of the SA in Sec.~\ref{sec:SA} and bound $\AD$ and $\mathcal{I}(\kappa)$ in Sec.~\ref{sec:A_D} and Sec.~\ref{sec:PS_model}, respectively. A performance bound of SA is derived in Sec.~\ref{sec:SA_bound}. In Sec.~\ref{sec:1D_Ising} we show how $\mathcal{I}(\kappa)$ can be calculated analytically for a 1D Ising chain. A numerical verification of the tightness of the bound on a Sherrington-Kirkpatrick (SK) model,
a fully connected spin glass with Gaussian couplings~\cite{Kirkpatrick1983,Panchenko2013}, is given in Sec.~\ref{sec:SA_numerics}. Finally, in Sec.~\ref{sec:SA_mu}, we provide further analytical and numerical discussion on the relaxation behaviour of SA by using SK models.

\subsection{SA algorithm}
\label{sec:SA}
In SA, the system state is updated iteratively until it approximately reaches equilibrium.
The evolution of the state can be viewed as a discrete-time Markov chain~\cite{levin2017markov}.
Suppose that temperature decreases from $T_\text{i}$ to $T_\text{f}$ in $\kappa$ steps, i.e., $T_\text{i}=T_0, T_1, \dots, T_k, \dots, T_\kappa=T_\text{f}$.
At step $k$, let the probability of the system staying in state $i$ be $p_i(k)$.
For each iteration step $k$, the state changes are described by
\begin{equation}
    p_i(k+1) = \sum_jP_{ij}(k+1)p_j(k),
    \eqL{SA_evolution}
\end{equation}
where $P_{ij}(k+1) = \mathrm{Prob}\{X(k+1)=i|X(k) = j\}$ is the transition matrix with $X(k)$ denoting the system state at step $k$,
satisfying $\sum_i P_{ij}(k) = 1,\forall j$ and $P_{ij}(k) \ge 0, \forall i,j$.

In typical SA algorithms, $P_{ij}(k)$ has the following form
\begin{equation}
    P_{ij}(k) = \left\{
        \begin{aligned}
            &\dfrac{1}{|\nbr{j}|} A_{ij}(k), & i \in \nbr{j}\\
            &1-\sum_{l(\neq j)}P_{lj}(k),  & i=j\\
            &0, & \text{otherwise}
        \end{aligned}
    \right.,
    \eqL{P(k)}
\end{equation}
where $|\nbr{j}|$ is the size of $\nbr{j}$, the neighbourhood of the state $j$ in which all states can be reached in one step from $j$, and the acceptance rate $A_{ij}(k)$ is the probability of accepting such a transition from $j$ to $i$. 

In line with SA simulating a thermalisation process, the detailed balance condition [\eqR{\ref{eq:detailed_balance}}] is imposed. Here, we more specifically  
adopt the acceptance rate from the {\em Glauber dynamics} which is widely used in SA and spin glass studies~\cite{Glauber1963,Levin2008,Walter2015,levin2017markov}, 
\begin{equation}
     A_{ij}(k) = \frac{1}{1+\exp(\beta(k)\Delta E_{ij})} = \frac{\gamma_i(k)}{\gamma_i(k) + \gamma_j(k)}, 
    \eqL{A_ij}
\end{equation}
where the energy difference $\Delta E_{ij} = E_i - E_j$.

\subsection{Bounding the iteration-averaged activity $\AD$ for SA}
\label{sec:A_D}
We now proceed to bound the activity that appears in \eqR{\ref{eq:main_bound}} for SA with Glauber dynamics. We first define the discrete-time activity $\mathcal{A}^\mathrm{D}$. We then derive an expression for $\mathcal{A}^\mathrm{D}$ under Glauber dynamics SA. We proceed to conjecture an upper bound on $\mathcal{A}^\mathrm{D}$ based on analytical and numerical evidence.  

Since the time evolution in SA [\eqR{\ref{eq:SA_evolution}}] is discrete, a discrete version of the activity is needed. We shall use a superscript $\mathrm{D}$ to denote {\em discrete}, and write the discrete-time activity as $\mathcal{A}^\mathrm{D}$.
In Appendix~\ref{app:C_to_D}, we show that any discrete-time Markov chain in \eqR{\ref{eq:SA_evolution}} can always be simulated by a continuous-time master equation in \eqR{\ref{eq:master_equation}}.
It is, therefore, natural to define the discrete-time activity as
\begin{equation}
    \mathcal{A}^\mathrm{D}(k) = \sum_i\sum_{j(\neq i)} P_{ij}(k+1) p_j(k).
    \label{eq:activity_D}
\end{equation}
which leads to the {\em iteration-averaged} activity
$\AD := \frac{1}{\kappa}\sum_{k=0}^{\kappa-1}\mathcal{A}^\mathrm{D}(k)$. $\AD$ equals the $\A$ of the corresponding continuous-time Markov chain (see Appendix~\ref{app:C_to_D}) and can be interpreted in terms of the expected number of jumps between micro-states. The RHS of \eqR{\ref{eq:activity_D}} is a sum over the probabilities of jumping from state $i$ to $j\neq i$, so we may write $\mathcal{A}^\mathrm{D}=p(\mathrm{jump})$. The expected number of jumps over the whole process, by inspection, is $\langle N_\mathrm{jumps}^\mathrm{D} \rangle=\sum_{k=0}^{\kappa-1} p(\mathrm{jump})$. Thus 
\begin{equation}
\langle N_\mathrm{jumps} ^\mathrm{D}\rangle = \AD\kappa.
\label{eq:Njumpsdiscrete}
\end{equation}
\eqR{\ref{eq:Njumpsdiscrete}} is the discrete-time analogue of \eqR{\ref{eq: expectN}}. For the case of continuous time, it is natural to define $\langle N_\mathrm{jumps} \rangle=\int_{t=0}^{\tau} \rho(\mathrm{jump}, t)dt$ where $\rho(\mathrm{jump}, t)$, which can be termed the rate of jumping, is the probability density for changing microstates in the infinitesimal time interval $dt$.

Substituting the general transition matrix $P_{ij}(k)$ of \eqR{\ref{eq:P(k)}} into Eq.~\eqref{eq:activity_D}, we have
\begin{equation}
    \mathcal{A}^\mathrm{D}(k) = \sum_i\sum_{j\in\nbr{i}}\dfrac{1}{|\nbr{j}|}A_{ij}(k+1)p_j(k).
    \eqL{A_D(k)}
\end{equation}
For simplicity, we denote $|\nbr{j}| = n_j$ and ignore the 
step dependence of quantities, writing $q(k) \equiv q$, where $q$ is some quantity. We use prime $'$ to denote the quantity at the next step: $q(k+1) \equiv q'$. 

Note that the acceptance rate $A_{ij}'$ defined in \eqR{\ref{eq:A_ij}} has the property: $A_{ij}' + A_{ji}' = 1$. We can therefore rewrite \eqR{\ref{eq:A_D(k)}} as
\begin{align}
    \mathcal{A}^\mathrm{D} &= \sum_i\sum_{j\in\nbr{i}}\dfrac{1}{n_j} A_{ij}'p_j \nonumber\\
    &= \dfrac{1}{2}\sum_i\sum_{j\in\nbr{i}}\left(A_{ij}'\dfrac{p_j}{n_j} + A_{ji}'\dfrac{p_i}{n_i}\right) \nonumber\\
    &= \dfrac{1}{2}\sum_i\sum_{j\in\nbr{i}} \dfrac{p_i}{n_i} + \dfrac{1}{2}\sum_i\sum_{j\in\nbr{i}}A_{ij}'\left(\dfrac{p_j}{n_j} - \dfrac{p_i}{n_i}\right) \nonumber\\
    &= \dfrac{1}{2} - \dfrac{1}{4}\sum_i\sum_{j\in\nbr{i}}\left(A_{ij}'-A_{ji}'\right)\left(\dfrac{p_i}{n_i} - \dfrac{p_j}{n_j}\right) \nonumber\\
    &= \dfrac{1}{2} - \dfrac{1}{4}\sum_i\sum_{j\in\nbr{i}}\dfrac{\gamma_i'-\gamma_j'}{\gamma_i'+\gamma_j'}\left(\dfrac{p_i}{n_i} - \dfrac{p_j}{n_j}\right).
\end{align}
Therefore, the activity, interpreted as the jumping probability, deviates from $1/2$, the unbiased case, by a number relevant to the order of occupation probabilities for neighbouring states. More precisely, we can consider the activity at the thermal state by setting 
$p_i = \gamma_i',\forall i$, which gives us
\begin{align}
        \mathcal{A}_\gamma^\mathrm{D} &\equiv \sum_i\sum_{j\in\nbr{i}}\dfrac{1}{n_j} A_{ij}'\gamma'_j \nonumber\\
        &= \dfrac{1}{2} - \dfrac{1}{4}\sum_i\sum_{j\in\nbr{i}}\dfrac{\gamma_i'-\gamma_j'}{\gamma_i'+\gamma_j'}\left(\dfrac{\gamma'_i}{n_i} - \dfrac{\gamma'_j}{n_j}\right).\label{eq:A_D_gamma_mid}
\end{align}
  The thermal state, at infinite temperature, becomes uniform (unbiased), such that  $\mathcal{A}^\mathrm{D}_\gamma = 1/2$, while, at zero temperature, $\mathcal{A}^\mathrm{D}_\gamma = 0$, meaning that any possible jumps are frozen. The analysis of the two extreme cases is intuitive and simple. However, the value of $\mathcal{A}^\mathrm{D}_\gamma$ at the intermediate temperatures highly depends on the neighbourhood structure of the energy landscape, making the determination of $\mathcal{A}_\gamma^\mathrm{D}$ quite complicated. 
  
 One common way to simplify the problem is by considering equal-sized neighbourhoods, i.~e.~
 $n_i= n, \forall i,$
 as in the Edwards-Anderson model~\cite{Edwards_1975} we shall introduce later.
Now, by \eqR{\ref{eq:A_D_gamma_mid}}, we have
\begin{equation}
    \mathcal{A}_\gamma^\mathrm{D} = \dfrac{1}{2} - \dfrac{1}{4n}\sum_i\sum_{j\in\nbr{i}}\dfrac{(\gamma_i'-\gamma_j')^2}{\gamma_i'+\gamma_j'}\le 1/2.
\end{equation}
We conclude that, at least for fixed neighbourhood size, $0\le\mathcal{A}_\gamma^\mathrm{D}\le 1/2$ during the annealing process where equality on both sides holds at zero and infinite temperatures, respectively. 

There is significant evidence that $\mathcal{A}^\mathrm{D}(k)$ also has a non-trivial upper bound in SA. In SA, the system, initially in a thermal state at some non-zero temperature,  keeps evolving toward a cooler thermal state. It is thus reasonable to conjecture that 
\begin{equation}
    \mathcal{A}^\mathrm{D}(k)\le1/2,\quad 0\le k\le \kappa.
    \label{eq:A_D_conjecture}
\end{equation}
Numerical evidence from simulations of a $7$-spin SK model (the formal definition of the SK model will be given in Sec.~\ref{sec:SA_numerics}),
is provided in \figR{\ref{fig:A_D}}. One sees \eqR{\ref{eq:A_D_conjecture}} is respected. Moreover, there is a decreasing trend of $\mathcal{A}^\mathrm{D}(k)$ during the annealing, which coincides with previous numerical results~\cite{Strenski1991} and could be a general feature of SA. As a consequence of the conjecture [\eqR{\ref{eq:A_D_conjecture}}], we have $\AD \le 1/2$.

\begin{figure}
    \centering
    \includegraphics[width=\linewidth]{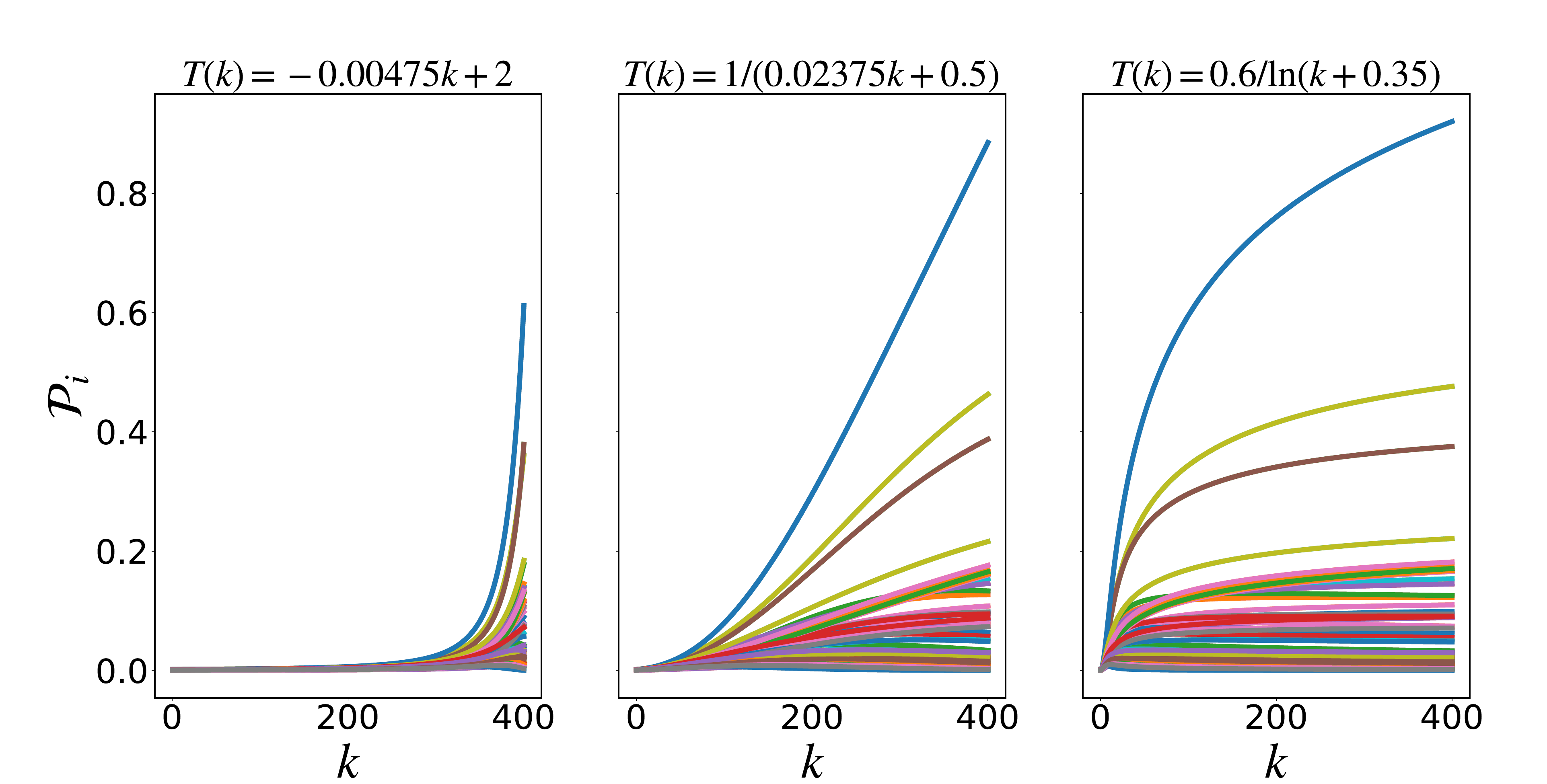}
    \caption{\textbf{Verification of the neighborhood-passivity assumption in a $7$-spin SK model.} The simulation takes $400$ steps with temperature changing from $T_\mathrm{i} = 2$ to $T_\mathrm{f} = 0.1$, according to the linear (left), inversely linear (middle) and inversely logarithmic (right) cooling schedules, respectively. Each curve represents the neighborhood passivity of a state in the state space. All curves are above $\mathcal{P}_i = 0$.}
    \label{fig:passivity}
\end{figure}

A further, intuitive, route to understanding  conjecture [\eqR{\ref{eq:A_D_conjecture}}] makes use of the tendency for lower energy states to have a higher probability in thermal scenarios. For  equal-sized neighborhoods, the activity $\mathcal{A}^\mathrm{D}$ is given by
\begin{equation}
    \mathcal{A}^\mathrm{D} = \dfrac{1}{2} - \dfrac{1}{4n}\sum_i\sum_{j\in\nbr{i}}\dfrac{(\gamma_i'-\gamma_j')(p_i-p_j)}{\gamma_i'+\gamma_j'}.
    \eqL{A_D_explicit}
\end{equation}
Adopting the concept of \textit{passive states} where the lower energy state has the higher probability of occupation~\cite{Pusz1978,Lenard1978,Marti2015,Sparaciari2017,Koukoulekidis2021},
we could assume the \textit{passivity in neighborhoods} and define the neighborhood-passivity $\mathcal{P}_i$ for state $i$,
\begin{align}
    \mathcal{P}_i &:= \sum_{j\in\nbr{i}}\dfrac{(\gamma_i'-\gamma_j')(p_i-p_j)}{\gamma_i'+\gamma_j'} \nonumber\\
    & = \sum_{j\in\nbr{i}}\tanh\left[\frac{\beta'}{2}(E_j-E_i)\right](p_i-p_j).
\end{align}
Note that if the two neighboring states $p_i$ and $p_j$ respect passivity, $\tanh\left[\beta'(E_j-E_i)/2\right](p_i-p_j)\ge 0$.
The passivity in neighborhoods thus requires $\mathcal{P}_i\ge 0, \forall i$. Once this assumption holds during the SA,
by \eqR{\ref{eq:A_D_explicit}}, we have $\mathcal{A}^\mathrm{D}\le 1/2$ for every step $k$. Consequently, $\AD \le 1/2$. This assumption can be justified by noticing that all thermal states are passive. Thus, as the system initially staying in a thermal state evolves towards a new thermal state in SA, we expect the passivity at least in the neighbourhoods to be preserved.
As is shown in \figR{\ref{fig:passivity}}, $\mathcal{P}_i\ge 0, \forall i$, for three different kinds of cooling schedules, which verifies our passivity assumption and gives
new numerical evidence for the conjecture of $\mathcal{A}^\mathrm{D}(k)\le1/2$, and hence $\AD \le 1/2$.
\begin{figure}
    \centering
    \includegraphics[width=\linewidth]{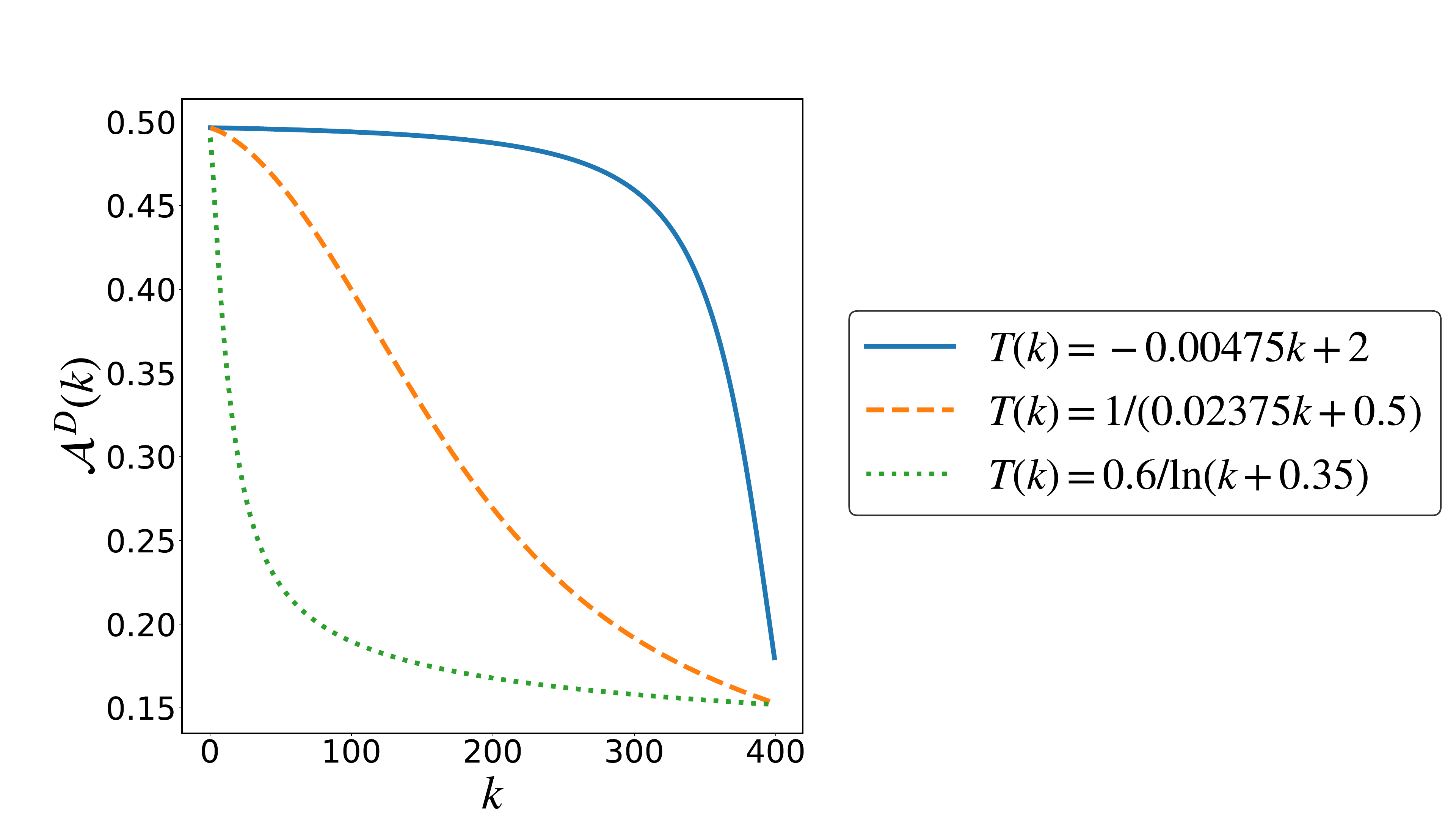}
    \caption{\textbf{The activity $\mathcal{A}^\mathrm{D}(k)$ in the SA of a $7$-spin SK model}. Simulations take $400$ steps with temperature changing from $T_\mathrm{i} = 2$ to $T_\mathrm{f} = 0.1$, according to the linear (blue solid line), inversely linear (orange dashed line) and inversely logarithmic (green dotted line) cooling schedules, respectively.}
    \figL{A_D}
\end{figure}

\subsection{Bounding the relative entropy change due to temperature variation $\mathcal{I}(\kappa)$}\label{sec:PS_model}
In this subsection, we will bound the relative entropy change from temperature variation, $\mathcal{I}$. In SA, $\mathcal{I}$ is evaluated by a discrete sum:
\begin{equation}
    \mathcal{I}(\kappa) = \sum_{k=1}^{\kappa}\left[E_p(k)-E_\gamma(k)\right]\left[\beta(k)-\beta(k-1)\right],
    \label{eq:I_kappa}
\end{equation}
which is shown to be equal to the $\mathcal{I}(\tau)$  of the underlying continuous-time process that simulates the discrete-time evolution in Appendix~\ref{app:C_to_D}.
Since $E_\gamma(k)$ is the energy expectation of the thermal state $\gamma$, although it apparently acquires the $k$-dependence from a cooling schedule $T(k)$, it is intrinsically a fixed function of temperature for a specific energy landscape. In this sense, we regard it as a problem-specific but dynamics-free quantity. In contrast, the average energy $E_p(k)$ requires information on the instantaneous system state $p(k)$. To access it, one has to solve the dynamics in \eqR{\ref{eq:SA_evolution}}, which is generally difficult and is the main obstacle to analyzing the performance of SA.
In the following, we will try to tackle it by providing an upper bound to $E_p(k)$ in terms of $E_\gamma(k)$. This is done via the use of a partial swap (PS) model, a simple thermalization protocol where the system state is swapped to the thermal state with a probability at each step~\cite{Scarani2002,Browne2014,Zhen2021}. 
We shall use the superscript PS to denote quantities in the partial swap model.  By showing that $E_p(k)\le E_p^\PS(k)$ for all $k$, we will derive a bound of $\mathcal{I}(\kappa)$ as $\mathcal{I}(\kappa)\le\mathcal{I}^\PS(\kappa)$.

The evolution of the state of the system $p^\mathrm{PS}$ in a partial swap model is given by
\begin{equation}
    p_i'^\PS = p_i^\PS + \mu^\PS(\gamma_i'^\PS - p_i^\PS),
    \eqL{PS_model}
\end{equation}
where $\mu^\PS$ is the so-called partial swap rate and the time dependence has been neglected and the prime $'$ denotes the next time step. Multiplying $E_i$ on both sides and summing over $i$ gives
\begin{equation}
    E_p'^\PS = E_p^\PS + \mu^\PS(E_\gamma'^\PS - E_p^\PS),
    \label{eq:E_PS_model}
\end{equation}
and therefore,
\begin{equation}
    \mu^\PS = \dfrac{E_p'^\PS-E_p^\PS}{E_\gamma'^\PS-E_p^\PS}.
\end{equation}
Similarly, we can define a \textit{relaxation rate} $\mu$ for the actual process in SA as
\begin{equation}
    \mu := \dfrac{E_p' - E_p}{E_\gamma' - E_p}.
    \label{eq:mu_def}
\end{equation}
This relaxation rate has been employed in analysing the minimal dissipation and designing the optimal cooling schedule in SA~\cite{Nulton1988,SALAMON1988423,Andresen1994}.
Note that by writing in this form, we require a positive $\mu$. Since in SA, $E_p\ge E_\gamma \ge E_\gamma'$, we need the assumption that $E_p \ge E_p'$. Such descending energy expectation has been witnessed in most uses of SA~\cite{Strenski1991,Ruppeiner1991,Anderson2002}.
$\mu^\PS$ and $\mu$ quantify the relaxation speeds of PS and the actual processes in terms of the energy differences, respectively. Intuitively, $\mu\ge\mu^\PS$ implies that the corresponding PS process
is \textit{slower} than the actual one in SA. This can be formulated by the following proposition.

\begin{proposition}
Consider that two identical systems initially in the same thermal state are annealed according to the same cooling schedule. Their relaxation dynamics are given by \eqR{\ref{eq:SA_evolution}} and \eqR{\ref{eq:PS_model}}, respectively. If $\mu\ge\mu^\PS$ during the annealing, $E_p\le E_p^\PS$ holds all the time.
\end{proposition}

\begin{proof}
Since the two processes correspond to the same energy landscape and cooling schedule, we have $E_\gamma'^\PS = E_\gamma'$. Let $\mu\ge\mu^\PS$, namely,
\begin{equation}
    \dfrac{E_p'-E_p}{E_\gamma'-E_p} \ge \dfrac{E_p'^\PS-E_p^\PS}{E_\gamma'-E_p^\PS}.
    \label{eq:compare_mu}
\end{equation}
As a partial swap rate, $\mu^\PS\le 1$. The insufficient thermalization leads to $E_p'^\PS \ge E_\gamma'$, and therefore, $\mu^\PS$ is a monotonically increasing function of $E_p^\PS$. If $E_p^\PS \ge E_p$,
\begin{equation}
    \dfrac{E_p'^\PS-E_p^\PS}{E_\gamma'-E_p^\PS} \ge \dfrac{E_p'^\PS-E_p}{E_\gamma'-E_p},
\end{equation}
and thus, by \eqR{\ref{eq:compare_mu}},
\begin{equation}
    \dfrac{E_p'-E_p}{E_\gamma'-E_p} \ge \dfrac{E_p'^\PS-E_p}{E_\gamma'-E_p}.
\end{equation}
In SA, $E_p \ge E_\gamma \ge E_\gamma'$. The above inequality gives us $E_p'^\PS \ge E_p'$. With the initial condition $E_p^\PS(0) = E_p(0)$, by induction, we proved that $\mu\ge\mu^\PS$ leads to $E_p^\PS \ge E_p$ all the time. 
\end{proof}

Knowing that $\mu\ge\mu^\PS$ does give us an upper bound on $E_p$, we further note that one can evaluate $E_p^\PS$ iteratively in \eqR{\ref{eq:E_PS_model}} by using $E_\gamma$ only. Replacing $E_p(k)$ with $E_p^\PS(k)$ in \eqR{\ref{eq:I_kappa}}, we thus find the desired bound $\mathcal{I}(\kappa)\le\mathcal{I}^\PS(\kappa)$.

Now the question is how to find such a $\mu^\PS$ less than $\mu$ when $\mu$ is not accessible (since its calculation still requires $E_p$). Recall that $\mu$ gauges the speed of the relaxation dynamics. The relaxation time $\tau_\mathrm{rel}(P) \equiv 1/(1-\lambda_2(P))$ of the discrete Markov chain~\cite{levin2017markov}, defined by the second largest eigenvalue of the transition matrix $P$, $\lambda_2(P)$, should be consistent with $\mu$. Therefore, we conjecture that the order of relaxation times of the two processes corresponding to $\mu$ and $\mu^\PS$ is the inverse order of $\mu$ and $\mu^\PS$, i.~e.~
\begin{equation}
    \tau_\mathrm{rel}\le\tau_\mathrm{rel}^\PS \Rightarrow \mu\ge\mu^\PS.
    \label{eq:ordering_conjecture}
\end{equation}
With this conjecture, we can compare $\lambda_2(P^\PS)$ and $\lambda_2(P)$ of the transition matrices $P^\PS$ and $P$ to find a condition on $\lambda_2(P^\PS)$. The desired $\mu^\PS$ can therefore be derived from $\lambda_2(P^\PS)$ by the following proposition.

\begin{proposition}
    For a partial swap model given by \eqR{\ref{eq:PS_model}}, $\lambda_2(P'^\PS) = 1-\mu^\PS$.
\end{proposition}

\begin{proof}
Let $N$ be the total number of states. The transition matrix $P'^\PS$ for such a PS model can be written explicitly as follow:
\begin{align}
        P'^{\mathrm{PS}} &=
    \begin{pmatrix}
        1-\mu+\mu\gamma'_1 & \mu\gamma'_1 & \cdots & \mu\gamma'_1 \\
        \mu\gamma'_2       &  1-\mu+\mu\gamma'_2 & \cdots & \mu\gamma'_2\\
        \vdots & \vdots & \ddots & \vdots \\
        \mu\gamma'_N & \mu\gamma'_N & \cdots & 1-\mu+\mu\gamma'_N\\
    \end{pmatrix} \nonumber\\
    &=
    (1-\mu)
    \begin{pmatrix}
        1 &  &  \\
          & \ddots &\\
        &  & 1\\
    \end{pmatrix}
    +\mu 
    \begin{pmatrix}
        \gamma'_1 & \cdots & \gamma'_1\\
        \gamma'_2 & \cdots & \gamma'_2\\
        \vdots &     & \vdots\\
        \gamma'_N & \cdots & \gamma'_N\\
    \end{pmatrix},
\end{align}
where we have neglected the superscript `$\PS$' of the matrix elements. Note that the second matrix on the right-hand side has a rank of $1$. By the rank-nullity theorem, for a matrix $T$, 
$\mathrm{Rank}(T) + \mathrm{Nullity}(T) = \mathrm{dim}(T)$, where $\mathrm{Nullity}(T) = \mathrm{dim}\{\vec{v}|T\vec{v} = 0\}$, 
the nullity of the second matrix is $N-1$, i.~e.~it has the eigenvalue of $0$ with multiplicity of $N-1$ and the other eigenvalue is 
$1$ given by its trace. Therefore, the eigenvalues of $P'^{\mathrm{PS}}$ are $\lambda_1(P'^{\mathrm{PS}}) = 1$ 
and $\lambda_2(P'^{\mathrm{PS}}) = \dots = \lambda_N(P'^{\mathrm{PS}}) =1-\mu^\PS$. 
\end{proof}

For the transition matrix $P$ in SA [\eqR{\ref{eq:P(k)}}], the second largest eigenvalue $\lambda_2(P)$ can be bounded as~\cite{Desai1993}
\begin{equation}
    \lambda_2(P) \le 1 - \dfrac{\omega_{\min}}{\gamma_{\max}}\lambda_2(Q),
    \eqL{lambda_2_P_bound}
\end{equation}
where $\omega_{\min} = \min_{i,j\in\nbr{i}} P_{ij}\gamma_j$ and $\gamma_{\max} = \max_i \gamma_i$. Here, $Q$ is the Laplacian matrix associated with the state space~\cite{Desai1993},
\begin{equation}
    Q_{ij} =\left\{
    \begin{aligned}
        &|\nbr{i}|\quad  & \text{if }i=j,\\
        &-1  & \text{if } j\in \nbr{i},\\
        &0   & \text{otherwise,}
    \end{aligned}
    \right.
\end{equation}
and $\lambda_2(Q)$ is the second smallest eigenvalue of $Q$. 
By \eqR{\ref{eq:P(k)}}, 
\begin{equation}
     \omega_{\min} = \min_{i,j\in \nbr{i}}\left(\frac{1}{|\nbr{j}|}\dfrac{\gamma_i\gamma_j}{\gamma_i+\gamma_j}\right)\ge \frac{1}{\mathcal{N}_{\max}}\dfrac{\gamma_{\min}}{2},
\end{equation}
where $\mathcal{N}_{\max} = \max_i |\nbr{i}|$ and $\gamma_{\min} = \min_i \gamma_i$.
Substituting the above into \eqR{\ref{eq:lambda_2_P_bound}}, we have
\begin{equation}
    \lambda_2(P) \le 1 - \dfrac{\lambda_2(Q)}{2\mathcal{N}_{\max}}\exp(-\beta\Delta E_{\max}).
\end{equation}
Since we want PS to be a slower process than SA, i.~e.~$\tau_\mathrm{rel}(P)\le\tau_\mathrm{rel}(P^\PS)$ followed by $\lambda_2(P)\le\lambda_2(P^\PS)$, we can simply set $\lambda_2(P'^\PS) = 1 - \mu^\PS = 1 - \dfrac{\lambda_2(Q)}{2\mathcal{N}_{\max}}\exp(-\beta'\Delta E_{\max})$, which gives us
\begin{equation}
    \mu^\PS(k) =  \dfrac{\lambda_2(Q)}{2\mathcal{N}_{\max}}\exp[-\beta(k+1)\Delta E_{\max}],
    \label{eq:mu_PS_general}
\end{equation}
where we have recovered the $k$-dependence for further uses. By the conjecture [\eqR{\ref{eq:compare_mu}}], such a $\mu^\PS(k)$ guarantees $\mu(k)\ge\mu^\PS(k)$. Hence, it leads to $\mathcal{I}(\kappa)\le\mathcal{I}^\PS(\kappa)$ as we discussed.

For a concrete example of the $\mu^\PS(k)$ of \eqR{\ref{eq:mu_PS_general}}, consider an $n$-spin Edwards-Anderson (EA) model~\cite{Edwards_1975}, a generalisation of the SK model with the Hamiltonian 
\begin{equation}
    H_{\mathrm{EA}} = \sum_{\langle k,l\rangle}g_{kl}s_ks_l,
    \label{eq:EA_Hamiltonian}
\end{equation}
where each spin $s_k\in\{-1,+1\}$ and $\langle k,l \rangle$ denotes the set of pairs of spins $s_k$ and $s_l$ having non-vanishing couplings $g_{kl}$. Here, we do not consider the self-energy terms, i.~e.~$g_{kk} = 0, \forall k$.
The system state $i$ is a vector representing the spin configuration $\vec{s}^{(i)} = \{s_1^{(i)},\dots,s_n^{(i)}\}$. The neighbors of $i$ are configurations with only one spin orienting oppositely with respect to $\vec{s}^{(i)}$. In other words, every neighborhood has the same size, i.~e.~$\mathcal{N}_{\max} = n$. Given that the total number of states is $2^n$, the state space is an $n$-dimensional hypercube, whose $\lambda_2(Q) = 2$~\cite{Desai1993}. In this case, \eqR{\ref{eq:mu_PS_general}} becomes
\begin{equation}
    \mu^\PS(k) =  \dfrac{1}{n}\exp[-\beta(k+1)\Delta E_{\max}].
    \label{eq:mu_PS_SK}
\end{equation}
We will use this $\mu^\PS(k)$ to bound our numerical results in Sec.~\ref{sec:SA_numerics}.

\subsection{Performance bound for SA}
\label{sec:SA_bound}
We now derive the performance bound for the SA we considered in this section. 

We define the entropy production in the discrete-time Markov chain, denoted by $\Sigma(\kappa)$, as the entropy production in the corresponding continuous-time Markov chain. The detailed justification is given in Appendix~\ref{app:C_to_D}, with the key results listed below. The discrete-time analogue of the speed limit in \eqR{\ref{eq:speed_limit}} is given by 
\begin{equation}
    \Sigma(\kappa) \ge \dfrac{(L_0-L_\kappa)^2}{2\AD\kappa}.
\end{equation}
Accordingly, \eqR{\ref{eq:UQineq}} becomes
\begin{equation}
    L_\kappa^2 \le -\frac{(L_0 - L_\kappa)^2}{\langle \mathcal{A}^\mathrm{D}\rangle_\kappa\kappa} + 2\mathcal{I}(\kappa),
\end{equation}
Substituting the conjectured bounds $\AD\le 1/2$ and $\mathcal{I}(\kappa)\le\mathcal{I}^\PS(\kappa)$ with the partial swap rate $\mu^\PS(k)$ in \eqR{\ref{eq:mu_PS_general}}, we have
\begin{equation}
    L_\kappa^2 \le - \dfrac{2(L_0-L_\kappa)^2}{\kappa} + 2\mathcal{I}^\PS(\kappa).
\end{equation}
Solving this inequality for $L_\kappa$, we have the performance bound of SA,
\begin{equation}
    L_\kappa \le \dfrac{2L_0 + \sqrt{2\kappa\left(-L_0^2 + \mathcal{I}^\PS(\kappa)\left(\kappa + 2\right)\right)}}{\kappa+2}.
    \label{eq:SA_bound}
\end{equation}
Compared with the discrete-time analogy of the bound \eqR{\ref{eq:main_bound}} (see Appendix~\ref{app:C_to_D}),
\begin{equation}
     L_\kappa \le \dfrac{L_0 + \sqrt{\langle \mathcal{A}^\mathrm{D}\rangle_\kappa\kappa\left(-L_0^2 + 2\mathcal{I}(\kappa)\left(\langle \mathcal{A}^\mathrm{D}\rangle_\kappa\kappa + 1\right)\right)}}{\langle \mathcal{A}^\mathrm{D}\rangle_\kappa\kappa+1},
     \label{eq:main_bound_discrete}
\end{equation}
\eqR{\ref{eq:SA_bound}} replaces the relative entropy term $\mathcal{I}$ with $\mathcal{I}^\PS$ associated with a partial swap model. Evaluating $\mathcal{I}^\PS(\kappa)$ does not require solving the evolution of $p(k)$ [\eqR{\ref{eq:SA_evolution}}] but only the information on $E_\gamma(k)$, which can be obtained by knowing the energy spectrum of the system.

For specific models, the energy spectrum or the probability distribution of energy levels may be known, from which one can construct the partition function $Z$ and find $E_\gamma$ by $E_\gamma = -\partial \ln Z / \partial \beta$.

For systems with {\em a priori} unknown energy levels, one can sample the energy landscape to estimate the spectrum. The computational resources required in evaluating the bound on the error of the solution are also then in fact much lower than the resources required {\em for evaluating the error} $L_\kappa$ via SA. To find $L_\kappa$, the distance between the statistical state $p(\kappa)$ and the equilibrium statistical state $\gamma(\kappa)$, directly through SA, many trajectories are indeed required, in order to obtain $p(k)$ at the $k$th step. A key advantage of \eqR{\ref{eq:SA_bound}}, is to replace the need to evaluate that distance with the need to know the energy spectrum. Given a state space of size $N$, one can find the energy landscape by the order of $N$ calls of the energy function. In contrast, for a faithful estimation of $p(k)$ via SA, the number of trajectories $N_{\mathrm{traj}}\gg N$ and each Monte Carlo time step for all trajectories needs $N_{\mathrm{traj}}$ calls of the energy function. Therefore, quantifying the computational cost by the number of calls of the energy function, evaluating our \eqR{\ref{eq:SA_bound}} is computationally much cheaper than performing SA to evaluate the error. 

In the following, we exemplify the calculation of the bound for 1D Ising chains with known energy spectrum in Sec.~\ref{sec:1D_Ising}, and study the tightness of the bound by numerical simulation on SK models in Sec.~\ref{sec:SA_numerics}.

\subsection{Example of 1D Ising Chain}
\label{sec:1D_Ising}
In order to show how this bound can be analytically calculated, we consider the SA on a 1D $n$-spin Ising chain, which is a simplified EA model, with the Hamiltonian given by~\cite{Burovski2019}
\begin{equation}
    H_{\mathrm{1D}} = -J\sum_{i=1}^{n}s_{i}s_{i+1},
\end{equation}
where $s_i\in\{-1,+1\}$, $J$ is the coupling constant and the periodic boundary condition is applied, i.~e.~$s_{n}s_{n+1} \equiv s_ns_1$.
For even $n$, the partition function $Z_\mathrm{1D}$ at the inverse temperature $\beta$ can be calculated as~\cite{Burovski2019}
\begin{equation}
    Z_\mathrm{1D} = 2^n \left[\cosh\left(\beta J\right)^n+\sinh\left(\beta J\right)^n\right].
\end{equation}
The equilibrium energy $E_{\gamma,\mathrm{1D}}$ is therefore
\begin{widetext}
    \begin{equation}
    \begin{aligned}
        E_{\gamma,\mathrm{1D}} &= -\frac{\partial \ln Z_\mathrm{1D}}{\partial \beta}
        &= -\frac{Jn\cosh\left(\beta J\right)\sinh\left(\beta J\right)\left[\cosh\left(\beta J\right)^{n-2}+\sinh\left(\beta J\right)^{n-2}\right]}{\cosh\left(\beta J\right)^n+\sinh\left(\beta J\right)^n}.
    \end{aligned}
\end{equation}
\end{widetext}
Using the partial swap model [\eqR{\ref{eq:E_PS_model}}], one can calculate $E^\PS_{p,\mathrm{1D}}(k)\equiv E^\PS_{p,\mathrm{1D}}(\beta(k))$ iteratively from $E_{\gamma,\mathrm{1D}}(k)\equiv E_{\gamma,\mathrm{1D}}(\beta(k))$ and find $\mathcal{I}^\PS_\mathrm{1D}(\kappa)$ by \eqR{\ref{eq:I_kappa}}. On the other hand, given the energy landscape, the initial distance $L_0 = \sum_i|\gamma_i(0)-\gamma_i(\kappa)|$ is known. The bound \eqR{\ref{eq:SA_bound}} can thus be evaluated for this model.

Apart from evaluating \eqR{\ref{eq:SA_bound}} exactly, we can also deduce the scaling of $L_\kappa$ for large $\kappa$. We consider the normal cooling schedules where the rate of varying temperature decreases with $\kappa$. Therefore, when $\kappa$ gets large, $\mathrm{d}k\equiv 1$ and $\mathrm{d}E^\PS_{p,\mathrm{1D}}(k) \equiv E^\PS_{p,\mathrm{1D}}(k+1) - E^\PS_{p,\mathrm{1D}}(k)$ are relatively small. \eqR{\ref{eq:E_PS_model}} can be written as a differential equation:
\begin{equation}
    \dot{E}^\PS_{p,\mathrm{1D}}(k) = -\mu^\mathrm{PS}_{\mathrm{1D}}(k)\left[E^\PS_{p,\mathrm{1D}}(k) - E_{\gamma,\mathrm{1D}}(k)\right],
\end{equation}
where $\dot{x}(k) \equiv \mathrm{d} x / \D k$. This equation can be solved with the initial condition $E^\PS_{p,\mathrm{1D}}(0) = E_{\gamma,\mathrm{1D}}(0)$ by (See Supplementary Material of Ref.~\cite{short}) 
\begin{equation}
    E^\PS_{p,\mathrm{1D}}(k) = E_{\gamma,\mathrm{1D}}(k) -\int_0^k\nE^{-\int_s^k\mu^\mathrm{PS}_{\mathrm{1D}}(s')\D s'}\dot{E}_{\gamma,\mathrm{1D}}(s)\D s.
    \eqL{E_PS_1D}
\end{equation}
Accordingly, for large $\kappa$, $\mathcal{I}^\PS_\mathrm{1D}(\kappa)$ calculated by \eqR{\ref{eq:I_kappa}} can be written as an integral in the similar form as \eqR{\ref{eq:I_tau}},
\begin{equation}
    \begin{aligned}
      \mathcal{I}^\PS_\mathrm{1D}(\kappa) &= \int_0^\kappa \left[E^\PS_{p,\mathrm{1D}}(k) - E_{\gamma,\mathrm{1D}}(k)\right] \dot{\beta}(k)\D k\\
      &= -\int_0^\kappa \int_0^k\nE^{-\int_s^k\mu^\mathrm{PS}_{\mathrm{1D}}(s')\D s'}\dot{E}_{\gamma,\mathrm{1D}}(s)\dot{\beta}(k) \D s \D k,
    \end{aligned}
\end{equation}
where the second line is from the substitution of \eqR{\ref{eq:E_PS_1D}}. We note that 
\begin{equation}
    \begin{aligned}
        \dot{E}_\gamma &= \sum_i E_i \dot\gamma_i
        = \sum_i E_i(E_\gamma-E_i)\gamma_i\dot{\beta}
        = - \sigma(E)^2_\gamma\, \dot{\beta},
    \end{aligned}
\end{equation}
where the second equality used $\dot{\gamma}_i = (E_\gamma-E_i)\gamma_i\dot{\beta}$ [\eqR{\ref{eq:gamma_dot}}] and $\sigma(E)^2_\gamma\equiv -\sum_i E_i(E_\gamma-E_i)\gamma_i$ is the energy variance for the thermal state $\gamma$. $\mathcal{I}^\PS_\mathrm{1D}(\kappa)$ can therefore be written as
\begin{widetext}
    \begin{equation}
    \mathcal{I}^\PS_\mathrm{1D}(\kappa) = \int_0^\kappa \int_0^k\exp\!\left[-\int_s^k\mu^\mathrm{PS}_{\mathrm{1D}}(s')\D s'\right]\sigma_\mathrm{1D}(E)^2_\gamma(s)\dot{\beta}(s)\dot{\beta}(k) \D s \D k,
    \label{eq:I_PS_1D_1}
\end{equation}
where
\begin{equation}
    \sigma_\mathrm{1D}(E)^2_\gamma =\frac{\partial^2 \ln Z_{\mathrm{1D}}}{\partial \beta^2} = \frac{4J^2n\left[\cosh\left(\beta J\right)^{2n}\sinh\left(\beta J\right)^2 - \cosh\left(\beta J\right)^{2}\sinh\left(\beta J\right)^{2n}+(n-1)\left(\cosh\left(\beta J\right)\sinh\left(\beta J\right)\right)^n\right]}{\left[\cosh\left(\beta J\right)^n + \sinh\left(\beta J\right)^{n}\right]^2\sinh\left(2 \beta J\right)^2}.
\end{equation}
For large $\beta J$, $\cosh\left(\beta J\right)\simeq\sinh\left(\beta J\right)$ and we have 
\begin{equation}
    \sigma_\mathrm{1D}(E)^2_\gamma \simeq \frac{J^2 n(n-1)}{\sinh\left(2 \beta J\right)^2} \simeq 4 J^2 n(n-1)\nE^{-4\beta J}.
\end{equation}
Consequently, by \eqR{\ref{eq:I_PS_1D_1}},
\begin{equation}
    \begin{aligned}
        \mathcal{I}_{1D}^\PS(\kappa) 
        &\simeq 4 J^2 n(n-1)\int_0^\kappa\int_0^k \exp\!\left[-\int_s^k \frac{1}{n}\exp[-2\beta(s')nJ]\D s'\right]\nE^{-4\beta(s) J} \dot{\beta}(s)\dot{\beta}(k)\D s\D k.
    \end{aligned}
    \label{eq:I_PS_1D_2}
\end{equation}
\end{widetext}
where as we suggested in \eqR{\ref{eq:mu_PS_SK}}, $\mu^\mathrm{PS}_{\mathrm{1D}}(k)$ is chosen as $\mu^\mathrm{PS}_{\mathrm{1D}}(k) \simeq \exp[-2\beta(k)nJ]/n$ with $\Delta E_{\max} = 2nJ$. \eqR{\ref{eq:I_PS_1D_2}} hence provides a systematic way to calculate $\mathcal{I}_{1D}^\PS(\tau)$ for any annealing schedule $\beta(k)$.

For example, Consider the cooling schedule:
$\beta(k) = \beta_\mathrm{i} + (\beta_\mathrm{f} - \beta_\mathrm{i}) k/\kappa$,
and use the lowest partial swap rate $\bar{\mu}\equiv \exp[-2\beta_\mathrm{f}nJ]/n\le\mu^\mathrm{PS}_{\mathrm{1D}}(k), \forall k$, for simplicity. We then have
\begin{widetext}
    \begin{equation}
    \mathcal{I}_{1D}^\PS(\kappa) \lesssim \frac{4 J^2 n(n-1)\nE^{-4\beta_\mathrm{i}J}(\beta_\mathrm{f} - \beta_\mathrm{i})^2}{-4J(\beta_\mathrm{f}-\beta_\mathrm{i}) + \bar{\mu}\kappa}\left[\frac{1-\nE^{-4J(\beta_\mathrm{f}-\beta_\mathrm{i})}}{4J(\beta_\mathrm{f}-\beta_\mathrm{i})} + \frac{-1 + \nE^{-\bar{\mu}\kappa}}{\bar{\mu}\kappa}\right].
\end{equation}
\end{widetext}
Choosing $J = 1/n$ to fix the energy range $\Delta E_{\max} = 2$, we find $\mathcal{I}_{1D}^\PS(\kappa) \simeq \mathcal{O}(n\kappa^{-1})$. Substituting it into the bound \eqR{\ref{eq:SA_bound}}, the scaling of the error $L_\kappa$ is given as $L_\kappa \simeq \mathcal{O}(n^{1/2}\kappa^{-1/2})$. This result shows how the annealing performance for the 1D Ising chain is influenced by the system and protocol parameters, for this specific cooling schedule, which to our knowledge has not been derived before.

\subsection{Numerical results for SK models}
\label{sec:SA_numerics}
To verify the tightness of our bound in \eqR{\ref{eq:SA_bound}}, we adopt an $n$-spin SK model to conduct SA. The Hamiltonian of an $n$-spin SK model is defined as~\cite{Kirkpatrick1983,Panchenko2013}
\begin{equation}
    H_{\mathrm{SK}} = \dfrac{1}{n^{3/2}}\sum_{k=1}^{n}\sum_{l=1}^{n}g_{kl}s_ks_l,
\end{equation}
where $s_k\in\{-1,+1\}$ and $\{g_{kl}\}_{k,l=1}^{n}$ is a collection of independent and identically distributed standard Gaussian random variables. The normalization factor $1/n^{3/2}$ is chosen to ensure that the maximum energy has the order of magnitude of 1~\cite{Panchenko2013}. The merit of this choice is that one can use the famous result of the SK model that $\lim_{n\rightarrow\infty}\mathbb{E}(H_{\mathrm{SK},\max}) \approx 0.76$, where $\mathbb{E}(\cdot)$ denotes the Gaussian expectation~\cite{Parisi1980,Panchenko2014}. The partial swap rate in \eqR{\ref{eq:mu_PS_SK}} is also applicable to the SK model. We thus have a nice estimation of $\Delta E_{\max}$ in the expression of $\mu^\PS(k)$ in the thermodynamic limit $n\rightarrow\infty$, i.~e.~
\begin{equation}
    \lim_{n\rightarrow\infty}\mathbb{E}(\Delta E_{\max}) = \lim_{n\rightarrow\infty}2\mathbb{E}(H_{\mathrm{SK},\max}) \approx 1.52.
\end{equation}
This result with \eqR{\ref{eq:mu_PS_SK}} provides a faithful way to find an appropriate $\mu^\PS(k)$ only in terms of cooling schedules for an SK model with large $n$.

In \figR{\ref{fig:result_SK}}, we present the numerical results of SA conducted on a $7$-spin SK model, in which the temperature is inverse-linearly decreased from $T_\mathrm{i} = 1.8$ to $T_\mathrm{f} = 0.8$ in $\kappa$ steps. 
The left plot validates that both evolution history-dependent or -independent bounds [\eqR{\ref{eq:main_bound_discrete}} and \eqR{\ref{eq:SA_bound}}] provide upper limits for the performance error $L_\kappa$ of SA. To reveal the $\kappa$-dependence of the annealing-related quantity, the relative entropy from temperature descending $\mathcal{I}(\kappa)$, bounds from both directions are presented in the right plot, where $\mathcal{I}^\PS(\kappa)$ is calculated by the method suggested in Sec.~\ref{sec:PS_model} and the lower bound is given by the trade-off relation in \eqR{\ref{eq:trade_off}} with $\AD\le 1/2$. Notably, the upper and lower bounds of $\mathcal{I}(\kappa)$ are saturated for fast (small $\kappa$) and slow (large $\kappa$) cooling, respectively, demonstrating the tightness of the two bounds.

For comparison, a similar simulation is provided in the accompanying paper~\cite{short}, which adopts linear cooling schedules.
\begin{figure}[htb]
    \centering
    \includegraphics[width=\columnwidth]{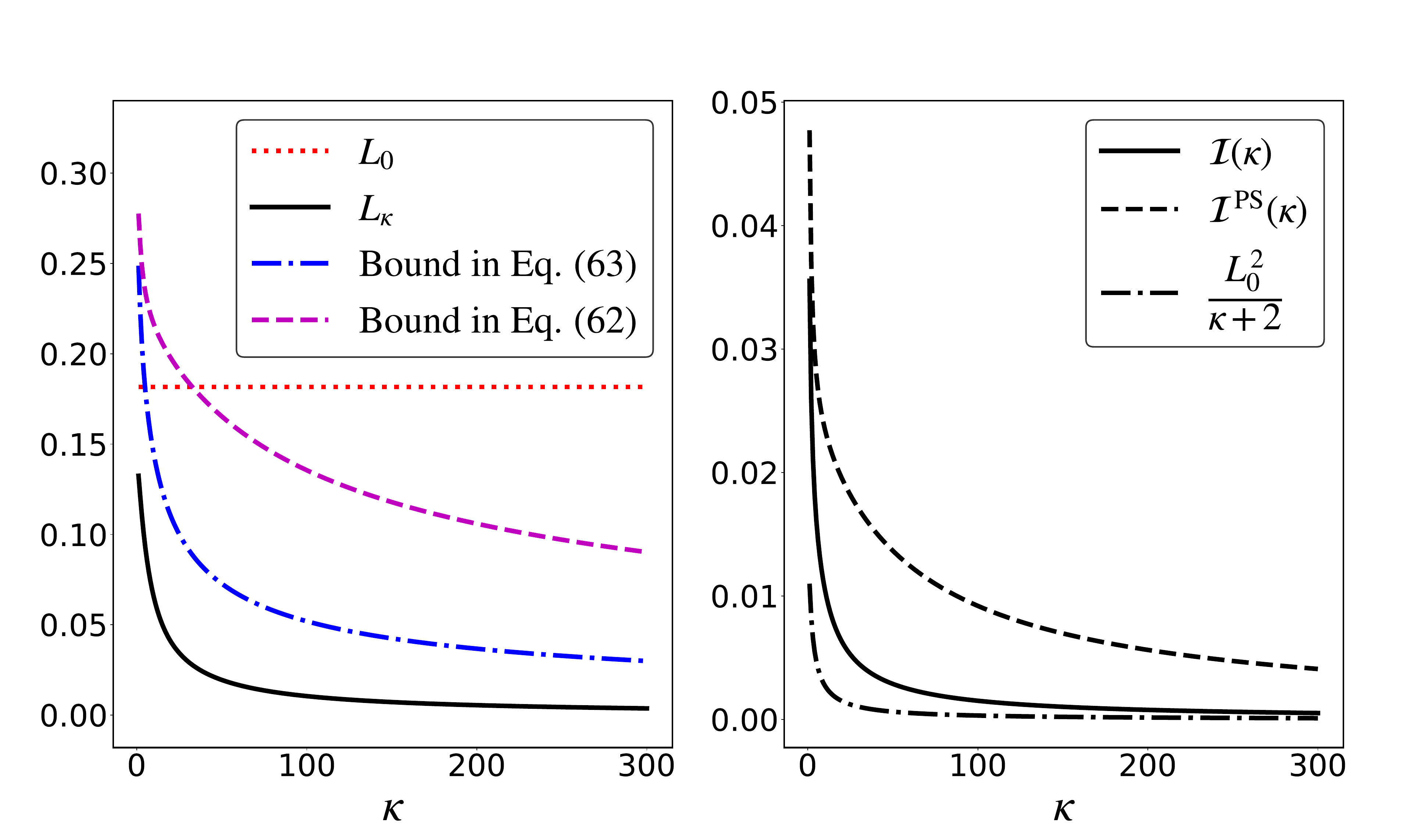}
    \caption{\figL{result_SK}{\bf Bounds on the performance of SA in solving a 7-spin SK model.} 
    $T_\mathrm{i} = 1.8$ is decreased inverse-linearly to $T_\mathrm{f} = 0.8$ in $\kappa$ steps. {\em Left:} The performance error $L_\kappa$ and our bounds in \eqR{\ref{eq:main_bound_discrete}} and in \eqR{\ref{eq:SA_bound}}, as well as the performance error of the initial state $L_0$.
    {\em Right:} The accumulated relative entropy due to temperature change $\mathcal{I}(\kappa)$ with our upper bound $\mathcal{I}^\PS(\kappa)$ and our lower bound from the trade-off relation between $\mathcal{I}(\kappa)$ and $\AD$ (for $\AD\le 1/2$).}
\end{figure}

\subsection{Relaxation rate $\mu$ of spin glasses in SA}
\label{sec:SA_mu}
From the previous discussion, we have seen that the choice of $\mu^\PS(k)$ in \eqR{\ref{eq:mu_PS_SK}} does give us an evolution history-independent bound for the example of an SK model. However, as is shown in \figR{\ref{fig:result_SK}}, there is still a gap between the history-independent bound in \eqR{\ref{eq:SA_bound}} and the history-dependent bound in \eqR{\ref{eq:main_bound_discrete}}.
A natural question to ask is whether \eqR{\ref{eq:mu_PS_SK}} is an optimal choice of $\mu^\PS(k)$ that satisfies $\mu(k)\ge\mu^\PS(k)$.
Can we find the largest $\mu^\PS(k)$ allowed?
In this section, we will pursue this possibility further by looking at the relaxation rate $\mu(k)$ defined in \eqR{\ref{eq:mu_def}} in more detail.

Referring to the time evolution of SA in \eqR{\ref{eq:SA_evolution}}, we have (again omitting the arguments $k$ and using $'$ to label quantities at $k+1$ step)
\begin{equation}
    p_i'= p_i + \dfrac{1}{n}\sum_{j\in\nbr{i}}\left(A_{ij}'p_j - A_{ji}'p_i\right),
\end{equation}
where we have considered the equal-sized neighborhoods, $|\nbr{i}| = n, \forall i$, and the acceptance rate $A_{ij}$ is given by \eqR{\ref{eq:A_ij}}. Thus, $E_p'$ is given by 
\begin{align}
        E_p' &= \sum_i E_i p_i' \nonumber\\
             &= E_p + \dfrac{1}{n}\sum_i\sum_{j\in\nbr{i}}\left(E_iA_{ij}'p_j - E_iA_{ji}'p_i\right) \nonumber\\
             &= E_p + \dfrac{1}{n}\sum_i\sum_{j\in\nbr{i}}\left(E_j-E_i\right)A_{ji}'p_i \nonumber\\
             &= E_p + \dfrac{1}{n}\sum_i\sum_{j\in\nbr{i}}\Delta E_{ji}\dfrac{\gamma_j' p_i}{\gamma_i' + \gamma_j'},
    \eqL{E_p_prime}
\end{align}
where we have defined $\Delta E_{ji} \equiv E_j-E_i$. Then
\begin{equation}
    E_p' - E_p = \dfrac{1}{n}\sum_i\sum_{j\in\nbr{i}}\Delta E_{ji}\dfrac{\gamma_j'}{\gamma_i'+\gamma_j'}(p_i-\gamma_i').
    \eqL{E_p_prime_E_p}
\end{equation}
To relate it with $\mu$, we want to rewrite the right-hand side of \eqR{\ref{eq:E_p_prime_E_p}} in terms of $E_\gamma' - E_p$. The following proposition is thus considered.

\begin{proposition}
    For an $n$-spin EA model with the Hamiltonian defined in \eqR{\ref{eq:EA_Hamiltonian}}, the following holds:
    \begin{equation}
    \sum_{j\in\nbr{i}}\Delta E_{ji} = - 4E_i.
    \eqL{sum_Delta_E}
    \end{equation}
\end{proposition}

\begin{proof}
We rewrite the Hamiltonian in \eqR{\ref{eq:EA_Hamiltonian}} as 
\begin{align}
    H_{\mathrm{EA}} &= \sum_{\langle k,l\rangle}g_{kl}s_ks_l \nonumber\\
    & = \dfrac{1}{2}\sum_{k=1}^{n}\sum_{l\in\mathcal{C}(k)}g_{kl}s_ks_l \nonumber\\
    & \equiv\dfrac{1}{2}\sum_{k=1}^{n}s_kh_k,
\end{align}
where $\mathcal{C}(k)$ is the set of spins interacting with spin $s_k$ and $h_k \equiv \sum_{l\in\mathcal{C}(k)}g_{kl}s_l$ is the effective field applied on $s_k$. The factor $1/2$ is from double-counting. Now consider a spin configuration (state $i$) $\vec{s}^{(i)} = \{s_1^{(i)},\dots,s_n^{(i)}\}$ with energy $E_i = \sum_{k=1}^{n} s_k^{(i)}h_k^{(i)} / 2$. A neighboring state $j$ can be generated by flipping the $j$-th spin in $\vec{s}^{(i)}$, i.~e.~$\vec{s}^{(j)} = \{s_1^{(i)},\dots,-s_j^{(i)}, \dots, s_n^{(i)}\}$. Therefore, the energy difference $\Delta E_{ji}$ is 
\begin{align}
    \Delta E_{ji} &= E_j - E_i \nonumber\\
    &= E_{\mathrm{rest}, j} - s_j^{(i)}h_j^{(i)} - \left(E_{\mathrm{rest}, j} + s_j^{(i)}h_j^{(i)}\right) \nonumber\\
    &= -2s_j^{(i)}h_j^{(i)},
\end{align}
where $E_{\mathrm{rest}, j}$ is the energy not involving the interactions with the spin $s_j$. Summing over all neighbors of $i$, we find that
\begin{equation}
    \sum_{j\in\nbr{i}}\Delta E_{ji} = \sum_{j=1}^{n} -2s_j^{(i)}h_j^{(i)} = - 4E_i.
\end{equation}
\end{proof}

As a result of this proposition, we have
\begin{equation}
    E_\gamma'-E_p = \dfrac{1}{4}\sum_i\sum_{j\in\nbr{i}}\Delta E_{ji}(p_i-\gamma_i').
\end{equation}
Therefore, \eqR{\ref{eq:E_p_prime_E_p}} can be written as
\begin{widetext}
\begin{align}
        E_p' - E_p &= \dfrac{1}{n}\sum_i\sum_{j\in\nbr{i}}\dfrac{\Delta E_{ji}}{2}(p_i-\gamma_i')  + \dfrac{1}{n}\sum_{i}\sum_{j\in\nbr{i}}\Delta E_{ji}\left(\dfrac{\gamma_j'}{\gamma_i'+\gamma_j'} - \dfrac{1}{2}\right)(p_i-\gamma_i')\nonumber\\
        &= \dfrac{2}{n}(E_\gamma'-E_p)  - \dfrac{1}{n}\sum_{i}\sum_{j\in\nbr{i}}\dfrac{\Delta E_{ji}}{2}\tanh\left(\beta'\dfrac{\Delta E_{ji}}{2}\right)(p_i-\gamma_i').
    \eqL{E_p_prime_E_p_final}
\end{align}
\end{widetext}
We then obtain
\begin{align}
        \mu &:= \dfrac{E_p' - E_p}{E_\gamma'-E_p} \nonumber\\
        &= \frac{2}{n} - \dfrac{1}{n}\sum_{i}\sum_{j\in\nbr{i}}\dfrac{\Delta E_{ji}}{2}\tanh\left(\beta'\dfrac{\Delta E_{ji}}{2}\right)\frac{p_i-\gamma_i'}{E_\gamma'-E_p}.
    \label{eq:mu_EA}
\end{align}
At high temperatures where $\beta'\Delta E_{\max}\ll 1$ with $\Delta E_{\max} = \max_iE_i - \min_iE_i$, $\mu\simeq 2/n$. This reflects the fact that in the high-temperature limit, the complex structure of the energy landscape can be neglected and the relaxation rate is only diminished by the size of the system.
\begin{figure}
    \centering
    \includegraphics[width=1.0\linewidth]{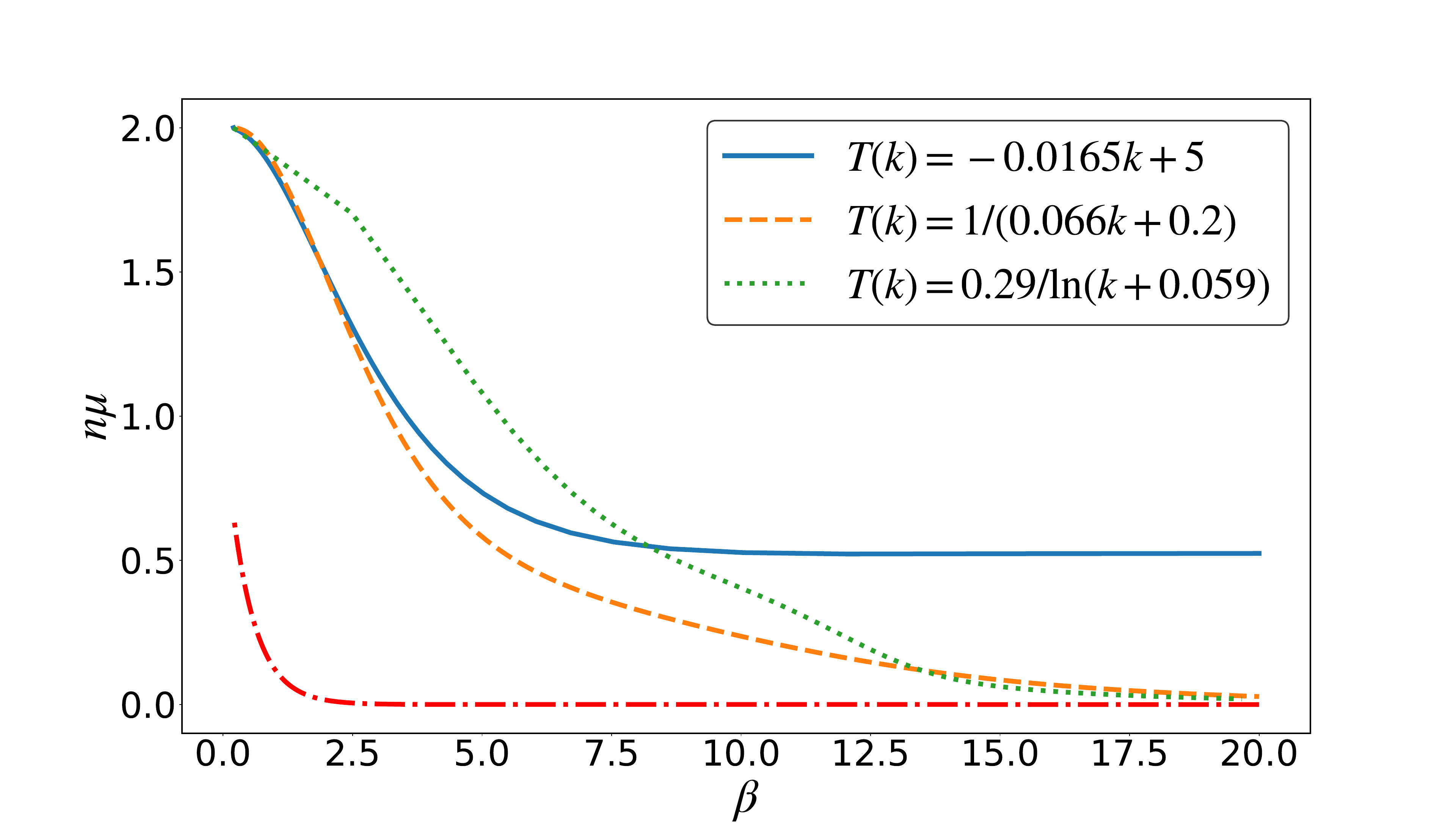}
    \caption{\textbf{The relaxation rate $\mu(\beta)$ (multiplied with $n$) in the SA of a $7$-spin SK model}. The simulation takes $300$ steps with temperature changing from $T_\mathrm{i} = 5$ to $T_\mathrm{f} = 0.05$, according to the linear (blue solid line), inversely linear (orange dashed line) and inversely logarithmic (green dotted line) cooling schedules, respectively. The partial swap rate $\mu^\PS(\beta)$ in \eqR{\ref{eq:SA_bound}} multiplied with $n$ is represented by the red dash-dotted line.}
    \figL{mu}
\end{figure}
As temperature decreases, the second term on the right-hand side of \eqR{\ref{eq:mu_EA}} becomes significant and therefore introduces more complication to the evaluation of $\mu$. We note that according to the cooling schedule, $\beta(k)$, $\mu(k(\beta)) \equiv \mu(\beta)$ can be viewed as a function of $\beta$. If $\mu(k)$ does not depend on the system state $p(k)$, $\mu(\beta)$ should be the same for all cooling schedules when $\beta$ is fixed. However, \figR{\ref{fig:mu}} provides a counter-example of a $7$-spin SK model showing that $\mu(k)$ is not a sole function of $\beta$ but relies on the history of $\beta$. Hence, finding $\mu(k)$ without referring to $p(k)$ seems intractable. Moreover, as is depicted in the red dash-dotted line in \figR{\ref{fig:mu}}, the $\mu^\PS$ in \eqR{\ref{eq:mu_PS_SK}} does bound $\mu$ irrespective of cooling schedules and is even saturated for large $\beta$. Therefore, at least for this specific example, at low temperatures, we cannot do much better than the $\mu^\PS$ in \eqR{\ref{eq:mu_PS_SK}}. On the other hand, \figR{\ref{fig:mu}} provides numerical evidence for our conjecture [\eqR{\ref{eq:ordering_conjecture}}].

Finally, we note that the asymptotic behaviour of $\mu\rightarrow 2/n$ at high temperatures, $\beta\rightarrow 0$, is recognized in \figR{\ref{fig:mu}} for all three cooling schedules. To take advantage of this property, one may use $\mu^\PS = 2/n$ for high temperatures and then use the $\mu^\PS$ in \eqR{\ref{eq:mu_PS_SK}} in further annealing to obtain a tighter bound than \eqR{\ref{eq:SA_bound}}.

\section{Conclusion}\label{sec:Conclusion}
In this article, we derived a general bound for quantifying the annealing performance. The bound is derived in the framework of finite-time stochastic thermodynamics and thus can be applied to nano-scale systems and particular quantum systems whose evolution satisfies stochastic master equations. The bound shows that the 1-norm distance between the final statistical state of the system and the final thermal state is restricted by two problem-specific parameters: the averaged activity $\A$ and the change of relative entropy due to the annealing $\mathcal{I}(\tau)$. The two quantities prove to obey a trade-off relation forced by the initial distance $L_0$.

We further applied this bound to study the finite-time performance of SA. Under a reasonable conjecture that the discrete activity $\mathcal{A}^\mathrm{D}\le1/2$ and purposing a slower partial swap model to bound $\mathcal{I}\le\mathcal{I}^\PS$, we obtained a modified bound which does not require the intermediate states of the system and only refers to the rule of temperature descent and the energy landscape. Consequently, the new bound applies to any Glauber dynamics SA schedules and any cost function. We employed 1D Ising chains as an example to illustrate how the bound can be evaluated analytically. To validate our conjectures, we presented numerical simulations of $7$-spin SK models, demonstrating the performance of our bounds. Above all, our results show that recent developments in stochastic thermodynamics can be adapted to characterize the finite time behaviour of SA, and we expect such approaches can be further developed to understand other Monte Carlo algorithms.

\vspace{5mm}
\begin{acknowledgments}
We gratefully acknowledge valuable discussions with Alexander Yosifov, Barry Sanders, Li Xiao, and Yu Chai. This work was supported by National Natural Science Foundation of China (Grants No. 12050410246, No. 12005091). 
\end{acknowledgments}

\appendix

\section{From continuous-time annealing to discrete-time SA}\appL{C_to_D}
Here we prove that the discrete-time evolution in SA, i.e.~\eqR{\ref{eq:SA_evolution}} in the main body, can always be simulated by a continuous-time master equation defined in Eq.~(\ref{eq:master_equation}) in the main body.
For the clarity of notations, below we write probabilities in the discrete-time with a superscript $\mathrm{D}$ and probabilities in the continuous time with a superscript $\mathrm{C}$.
Then, our aim is to prove that:
\begin{equation}
    p^\mathrm{D}_i(k+1) = \sum_{j}P_{ij}(k+1)p^\mathrm{D}_j(k),
    \eqL{pD_evolution}
\end{equation}
can be constructed by
\begin{equation}
    \dot{p}^\mathrm{C}_i(t) = \sum_j\Gamma_{ij}(t)p^\mathrm{C}_j(t),
    \eqL{pC_evolution}
\end{equation}
such that
\begin{equation}
    p^\mathrm{D}_i(k) = p^\mathrm{C}_i(t_k),
    \label{eq:equivalence_DC}
\end{equation}
for times $t_{0}=0,t_{1},\dots,t_{k},\dots,t_{\kappa}=\tau$ with $\kappa$ denoting the number of iterations in SA.

Integrating \eqR{\ref{eq:pC_evolution}} from time $t_k$ to 
time $t_{k+1}$, we obtain
\begin{equation}
    \begin{aligned}
        p_i^\mathrm{C}(t_{k+1}) &= p_i^\mathrm{C}(t_k) + \int_{t_k}^{t_{k+1}} \D t  \sum_j\Gamma_{ij}(t)p^\mathrm{C}_j(t)\\
        &= \sum_{j}W_{ij}(k+1)p_j^\mathrm{C}(t_k),
    \end{aligned}
    \eqL{W_evolution}
\end{equation}
where we have defined
\begin{equation}
    W_{ij}(k+1) = \delta_{ij} + \int_{t_k}^{t_{k+1}} \D t \,\Gamma_{ij}(t)p^\mathrm{C}_j(t)/p^\mathrm{C}_j(t_k),\eqL{W_ij}
\end{equation}
for each $k=0,1,\dots,\kappa-1$, where $\delta_{ij}$ is the Kronecker delta.
To show that there always exists a valid transition rate matrix making
\eqR{\ref{eq:W_evolution}} reproduce \eqR{\ref{eq:pD_evolution}}, we introduce
\begin{equation}
    \Gamma_{ij}(t) = \sum_{k=0}^{\kappa-1}\delta
    \left(t-t_k^+\right)
    P_{ij}(k+1),
    \eqL{Gamma_ij}
\end{equation}
for $i\neq j$ and $\Gamma_{ii}(t) = - \sum_{j(\neq i)}\Gamma_{ji}(t)$.
Here, $\delta(\cdot)$ is the Dirac $\delta$-function and 
$t_k^+>t_k$ is a value close to $t_k$.

As we will show, $P_{ij}(k+1)$ guarantees that the transition rate defined in \eqR{\ref{eq:Gamma_ij}} induces the required thermal states and detailed balance relation.
Due to the relation of $P_{ij}(k+1)$ demanded by SA
\begin{equation}
    P_{ij}(k+1)\gamma_j^\mathrm{D}(k+1) = P_{ji}(k+1)\gamma_i^\mathrm{D}(k+1), \eqL{P_ij-DB}
\end{equation}
it can be verified that the state defined as
\begin{equation}
\gamma^{\mathrm{C}}\left(t\right)=\gamma^{\mathrm{D}}\left(k+1\right),\text{ if }t_{k}^{+}<t\leqslant t_{k+1}
\label{eq:thermal_equivalence_DC}
\end{equation}
satisfies
\begin{equation}
    \Gamma_{ij}(t)\gamma_j^\mathrm{C}(t) = \Gamma_{ji}(t)\gamma_i^\mathrm{C}(t),
    \eqL{DB_Gamma}
\end{equation}
for all $i\neq j$.
Thus, $\gamma^{\mathrm{C}}\left(t\right)$ always satisfies the detailed balance condition and is a valid equilibrium state for the process governed by \eqR{\ref{eq:Gamma_ij}}.
Meanwhile, substituting \eqR{\ref{eq:Gamma_ij}} into \eqR{\ref{eq:W_ij}}, we obtain
\begin{equation}
    \begin{aligned}
        W_{ij}(k+1) &= P_{ij}(k+1), & \forall i\neq j,\\
        W_{ii}(k+1) &= P_{ii}(k+1), & \forall i.
    \end{aligned}
\end{equation}
where we have used the fact that $p_j^\mathrm{C}(t_k^+) / p_j^\mathrm{C}(t_k) \rightarrow 1$.

Finally, by taking $p^\mathrm{D}_i(0) = p^\mathrm{C}_i(t_0)$, we obtain that the continuous-time  evolution described by \eqR{\ref{eq:pC_evolution}} and \eqR{\ref{eq:Gamma_ij}} is in accordance with the discrete-time evolution described by
\eqR{\ref{eq:pD_evolution}} and \eqR{\ref{eq:P_ij-DB}} at discrete times $t_0,\dots,t_{\kappa}$.

Consequently, the time-averaged activity $\langle \mathcal{A}^\mathrm{C}\rangle_\tau$ is defined as
\begin{equation}
    \begin{aligned}
        \langle \mathcal{A}^\mathrm{C}\rangle_\tau &= \frac{1}{\tau}\int_0^\tau \D t\sum_{i}\sum_{j(\neq i)}\Gamma_{ij}(t)p_j^\mathrm{C}(t)\\
        &= \frac{1}{\tau}\sum_{k=0}^{\kappa-1}\int_{t_k}^{t_{k+1}}\D t\sum_{i}\sum_{j(\neq i)}\Gamma_{ij}(t)p_j^\mathrm{C}(t)\\
        &= \frac{1}{\tau}\sum_{k=0}^{\kappa-1}\sum_{i}\sum_{j(\neq i)}P_{ij}(k+1)p^\mathrm{D}_j(k).
    \end{aligned}
\end{equation}
To obtain an analogy of $\langle \mathcal{A}^\mathrm{C}\rangle_\tau$ for the discrete case in SA, we suppose that the time interval $t_{k+1} - t_{k}$ is a unit time $1$, such that $\tau = \kappa$, and define the \textit{iteration-averaged} activity $\langle \mathcal{A}^\mathrm{D}\rangle_\kappa$ as
\begin{align}
    \langle \mathcal{A}^\mathrm{D}\rangle_\kappa &= \langle \mathcal{A}^\mathrm{C}\rangle_\tau \\ &= \frac{1}{\kappa}\sum_{k=0}^{\kappa-1}\sum_{i}\sum_{j(\neq i)}P_{ij}(k+1)p^\mathrm{D}_j(k)\\ &\equiv \frac{1}{\kappa}\sum_{k=0}^{\kappa-1} \mathcal{A}^\mathrm{D}(k),
    \label{eq:A_equivalence_DC}
\end{align}
where the discrete activity is defined as
\begin{equation}
    \mathcal{A}^\mathrm{D}(k) = \sum_{i}\sum_{j(\neq i)}P_{ij}(k+1)p^\mathrm{D}_j(k).
\end{equation} 
Since the entropy production depends on the path of the statistical state of the system, it is more sensible to define the entropy production for the discrete-time evolution according to the underlying continuous-time evolution, in which the entropy production rate is well-defined as Eq.~(\ref{eq:entropy_prod_rate}) in the main body, i.~e.~
\begin{equation}
    \dot{\Sigma}^\mathrm{C} = \frac{1}{2}\sum_{i}\sum_{j(\neq i)}\left[\Gamma_{ij}p_j^\mathrm{C}-\Gamma_{ji}p_i^\mathrm{C}\right]\ln\frac{\Gamma_{ij}p_j^\mathrm{C}}{\Gamma_{ji}p_i^\mathrm{C}}.
\end{equation}
Therefore, we define
\begin{equation}
    \Sigma^\mathrm{D}(\kappa) \equiv \Sigma^\mathrm{C}(\tau) = \int_0^\tau \dot{\Sigma}^\mathrm{C}(t) \D t.
\end{equation}
By the `speed limit' with $\Sigma^\mathrm{C}(\tau)$ in Eq.~(\ref{eq:speed_limit}) in the main body, we have
\begin{equation}
    \Sigma^\mathrm{D}(\kappa) = \Sigma^\mathrm{C}(\tau) \ge \frac{(L_0 - L_\tau)^2}{2\langle \mathcal{A}^\mathrm{C}\rangle_\tau\tau} = \frac{(L_0 - L_\kappa)^2}{2\langle \mathcal{A}^\mathrm{D}\rangle_\kappa\kappa},
    \label{eq:speed_limit_D}
\end{equation}
where we have used $\tau = \kappa$, Eq.~(\ref{eq:A_equivalence_DC}) and
\begin{align}
    L_0 &= \sum_i|p^\mathrm{C}_i(0)-\gamma^\mathrm{C}_i(\tau)| = \sum_i|p^\mathrm{D}_i(0)-\gamma^\mathrm{D}_i(\kappa)|,\\
    L_\tau& = \sum_i|p^\mathrm{C}_i(\tau)-\gamma^\mathrm{C}_i(\tau)|,\\
    L_\kappa& = \sum_i|p^\mathrm{D}_i(\kappa)-\gamma^\mathrm{D}_i(\kappa)| = L_\tau.
\end{align}
The last equality $L_\kappa = L_\tau$ is from $p^\mathrm{D}(\kappa) = p^\mathrm{C}(\tau)$ by Eq.~(\ref{eq:equivalence_DC}) and $\gamma^\mathrm{D}(\kappa) = \gamma^\mathrm{C}(\tau)$ by Eq.~(\ref{eq:thermal_equivalence_DC}). Eq.~(\ref{eq:speed_limit_D}) is therefore a `speed limit' in the discrete-time case.

Accordingly, we can define the accumulated change in relative entropy $\mathcal{I}^\mathrm{D}(\kappa)$
by $\mathcal{I}^\mathrm{C}(\tau)$ in Eq.~(\ref{eq:I_tau}) in the main body, such that
\begin{align}
    \mathcal{I}^\mathrm{D}(\kappa) &\equiv \mathcal{I}^\mathrm{C}(\tau) \\
    &= \int_0^\tau \left[E_p^\mathrm{C}(t)-E_\gamma^\mathrm{C}(t)\right]\dot{\beta}(t)\mathrm{d} t\\
    &= \sum_i \int_0^\tau E_i\left[p_i^\mathrm{C}(t)-\gamma_i^\mathrm{C}(t)\right]\dot{\beta}(t)\mathrm{d} t\\
    &= \sum_{k=0}^{\kappa-1} \sum_i \int_{t_k}^{t_{k+1}} E_i\left[p_i^\mathrm{C}(t)-\gamma_i^\mathrm{C}(t)\right]\dot{\beta}(t)\mathrm{d} t \\
    &= \sum_{k=0}^{\kappa-1} \sum_i E_i \left[p_i^\mathrm{C}(t_{k+1})-\gamma_i^\mathrm{C}(t_{k+1})\right]\left[\beta(t_{k+1})-\beta(t_{k})\right]\\
    &= \sum_{k=0}^{\kappa-1} \sum_i E_i \left[p_i^\mathrm{D}(k+1)-\gamma_i^\mathrm{D}(k+1)\right]\left[\beta(k+1)-\beta(k)\right]\\
    &= \sum_{k=0}^{\kappa-1}\left[E_p^\mathrm{D}(k+1)-E_\gamma^\mathrm{D}(k+1)\right]\left[\beta(k+1)-\beta(k)\right],
\end{align}
where in the 4th line we used the fact that $p^\mathrm{C}(t) = p^\mathrm{C}(t_{k+1})$ and $\gamma^\mathrm{C}(t) = \gamma^\mathrm{C}(t_{k+1})$ for $t_k^+ < t \le t_{k+1}$, while in the 5th line, $\beta(t_k) = \beta(k)$ is in accordance with Eq.~(\ref{eq:thermal_equivalence_DC}).

From Eq.~(\ref{eq:relative_entropy}) in the main body, we have
\begin{align}
    S(p^\mathrm{D}(\kappa)||\gamma^\mathrm{D}(\kappa)) &= S(p^\mathrm{C}(\tau)||\gamma^\mathrm{C}(\tau)) \\&= -\Sigma^\mathrm{C}(\tau) + \mathcal{I}^\mathrm{C}(\tau) \\&= -\Sigma^\mathrm{D}(\kappa) + \mathcal{I}^\mathrm{D}(\kappa).
\end{align}
Using Pinsker's inequality, $L_\kappa^2\le 2 S(p^\mathrm{D}(\kappa)||\gamma^\mathrm{D}(\kappa))$ and the `speed limit' [Eq.~(\ref{eq:speed_limit_D})], we obtain
\begin{equation}
    L_\kappa^2 \le -\frac{(L_0 - L_\kappa)^2}{\langle \mathcal{A}^\mathrm{D}\rangle_\kappa\kappa} + 2\mathcal{I}^\mathrm{D}(\kappa),
\end{equation}
which is an analogy of Eq.~(\ref{eq:UQineq}) in the main body. Solving the inequality for $L_\kappa$ gives us the discrete-time performance bound:
\begin{equation}
     L_\kappa \le \dfrac{L_0 + \sqrt{\langle \mathcal{A}^\mathrm{D}\rangle_\kappa\kappa\left(-L_0^2 + 2\mathcal{I}^\mathrm{D}(\kappa)\left(\langle \mathcal{A}^\mathrm{D}\rangle_\kappa\kappa + 1\right)\right)}}{\langle \mathcal{A}^\mathrm{D}\rangle_\kappa\kappa+1},
\end{equation}
analogue to Eq.~(\ref{eq:main_bound}) in the main body. 

\bibliography{references}

\end{document}